\documentclass[sigconf]{acmart}

\AtBeginDocument{%
  \providecommand\BibTeX{{%
    \normalfont B\kern-0.5em{\scshape i\kern-0.25em b}\kern-0.8em\TeX}}}

\copyrightyear{2020}
\acmYear{2020}
\setcopyright{acmcopyright}
\acmConference[FPGA '20]{Proceedings of the 2020 ACM/SIGDA International Symposium on Field-Programmable Gate Arrays}{February 23--25, 2020}{Seaside, CA, USA}
\acmBooktitle{Proceedings of the 2020 ACM/SIGDA International Symposium on Field-Programmable Gate Arrays (FPGA '20), February 23--25, 2020, Seaside, CA, USA}
\acmPrice{15.00}
\acmDOI{10.1145/3373087.3375303}
\acmISBN{978-1-4503-7099-8/20/02}

\usepackage{multirow}
\usepackage{pifont}
\usepackage{subfig}
\usepackage{amsmath}
\usepackage{tabularx} 
\usepackage{mathtools}
\usepackage{bm}

\DeclarePairedDelimiter\floor{\lfloor}{\rfloor}
\usepackage{tikz}

\newcommand{\PLH}{{\mkern-2mu\times\mkern-2mu}}
\newsavebox{\measurebox}
\usepackage{soul}
\usepackage{amssymb}
\newcommand\numberthis{\addtocounter{equation}{1}\tag{\theequation}}
\usepackage{bbding}
\definecolor{amber}{rgb}{1.0, 0.75, 0.0}

\begin{document}

\title{LUXOR: An FPGA Logic Cell Architecture for Efficient Compressor Tree Implementations}

\author{SeyedRamin Rasoulinezhad$^1$, Siddhartha$^1$, Hao Zhou$^2$, Lingli Wang$^2$, David Boland$^1$, Philip H.W. Leong$^1$}
\affiliation{
   \institution{$^1$School of Electrical and Information Engineering, The University of Sydney, Sydney, 2006, Australia   
   \\$^2$State Key Lab of ASIC and System, Fudan University, Shanghai 201203, China \\
   \{seyedramin.rasoulinezhad, siddhartha.siddhartha, david.boland, philip.leong\}@sydney.edu.au\\ 
   \{zhouhao, llwang\}@fudan.edu.cn}
}
 


\begin{abstract}

We propose two tiers of modifications to FPGA logic cell architecture to deliver a variety of performance and utilization benefits with only minor area overheads. In the first tier, we augment existing commercial logic cell datapaths with a 6-input XOR gate in order to improve the expressiveness of each element, while maintaining backward compatibility. This new architecture is vendor-agnostic, and we refer to it as LUXOR. We also consider a secondary tier of vendor-specific modifications to both Xilinx and Intel FPGAs, which we refer to as X-LUXOR+ and I-LUXOR+ respectively. We demonstrate that compressor tree synthesis using generalized parallel counters (GPCs) is further improved with the proposed modifications.
Using both the Intel adaptive logic module and the Xilinx slice at the 65nm technology node for a comparative study, it is shown that the silicon area overhead is less than 0.5\% for LUXOR and 5--6\% for LUXOR+, while the delay increments are 1--6\% and 3--9\% respectively. We demonstrate that LUXOR can deliver an average reduction of 13--19\% in logic utilization on micro-benchmarks from a variety of domains.
BNN benchmarks benefit the most with an average reduction of 37--47\% in logic utilization, which is due to the highly-efficient mapping of the XnorPopcount operation on our proposed LUXOR+ logic cells.


\end{abstract}

\maketitle


\section{Introduction}

The design of parallel computer arithmetic circuits is a well established field of research dating back to the works of Wallace~\cite{wallace1964suggestion}, Dadda~\cite{dadda1965some}, Swartzlander~\cite{swartzlander1973parallel}, Verma~\cite{verma2008data}, and others. In the context of field-programmable gate arrays (FPGAs), there has always been interest in specialized arithmetic primitives which improve performance over a wide range of application domains. One such primitive, Generalized Parallel Counters (GPCs), enables fast accumulation of compressor trees. Work from Parandeh-Afshar~et.~al.~\cite{parandeh2009improving} motivated the use of GPCs on FPGAs, while Kumm~et.~al.~\cite{DBLP:conf/fpl/KummZ14} demonstrated software techniques that automate the design of optimal compressor tree implementations for FPGAs. However, modern FPGA lookup table (LUT) based architectures are not particularly efficient for implementation of compressor trees~\cite{DBLP:journals/trets/Parandeh-AfsharBI09}.


In this paper, we show that support for compressor trees in FPGAs could be significantly improved through minor modifications to the logic element (LE). This is beneficial for implementing low-precision and multi-operand operations. One example of interest is that compressor trees and GPCs can be used to accelerate the XnorPopcount operations within binarized neural networks (BNNs)~\cite{DBLP:journals/corr/abs-1809-04570}, 
which forms the critical path of the model's execution. BNNs enable neural networks to be utilized in resource constrained applications and can be deployed efficiently on FPGAs~\cite{,Zhao:2017:ABC:3020078.3021741,liang2018fp}; our optimizations would improve their performance further.

LUXOR is a portmanteau of the acronyms LUT and XOR. Its design is motivated by the observation that the Boolean XOR operation is very commonly found in optimized compressor trees. This is corroborated by Verma~et.~al. in~\cite{DBLP:conf/date/VermaI07}, where they exploited the correlations between the operands of the XOR function to improve delay for ASIC implementations. Our goal is to utilize this insight in a similar vein, but optimized for FPGAs.    

Our proposed changes provide a means to efficiently implement compressor trees using new area-optimized GPCs, which can all be applied to a large variety and/or important classes of applications. The contributions of this paper can be summarized as follows:
\begin{itemize}
    \item A new logic element, LUXOR, that integrates a 6-input XOR gate with commercial FPGA logic elements. This architecture independent modification improves the implementation of XnorPopcount operation and the most commonly used GPC.
    
    \item LUXOR+, an amalgamation of LUXOR with further Intel (I-LUXOR+) and Xilinx (X-LUXOR+) architecture-specific optimizations to achieve further resource reduction. To the best of our knowledge, this leads to the most efficient reported logic element based GPC, called C06060606, which can be mapped to just a single Xilinx slice.
    
    \item A novel integer linear programming (ILP) formulation based on the flexible Ternary Adder approach proposed in \cite{DBLP:journals/corr/abs-1806-08095} to optimally map compressor tree problems to LUXOR cells.

    \item Quantitative investigation of the benefits of LUXOR and LUXOR+ architectures using a set of more than 50 micro-benchmarks. 
    Our results also show the positive benefits of the proposed LUXOR and LUXOR+ enhancements in SMIC 65nm standard cell technology. 
    
    \item The ILP-based compressor tree synthesizer, benchmarks and design files required to generate the results in this paper are open source to support reproducible research, and available at 
    \texttt{github.com/raminrasoulinezhad/LUXOR\_FPGA20}.
\end{itemize}

The remainder of the paper is organized as follows. In Section~\ref{se:background}, we provide background on parallel counters, GPCs, compressors, and compressor trees. Our LUXOR and LUXOR+ enhancements are presented in Section~\ref{se:luxor}, and the accompanying ILP formulation in Section~\ref{se:ilp}. 
The experiment results are given in Section~\ref{se:results}. Finally, we present conclusions in Section~\ref{se:conclusion}.


\section{Background}
\label{se:background}

\subsection{Parallel Counters}

\begin{figure}[h]
    \centering
    \subfloat[][FA takes in three bits (dots) and produces two outputs: sum and carry] {
        \centering
        \includegraphics[width=0.35\linewidth]{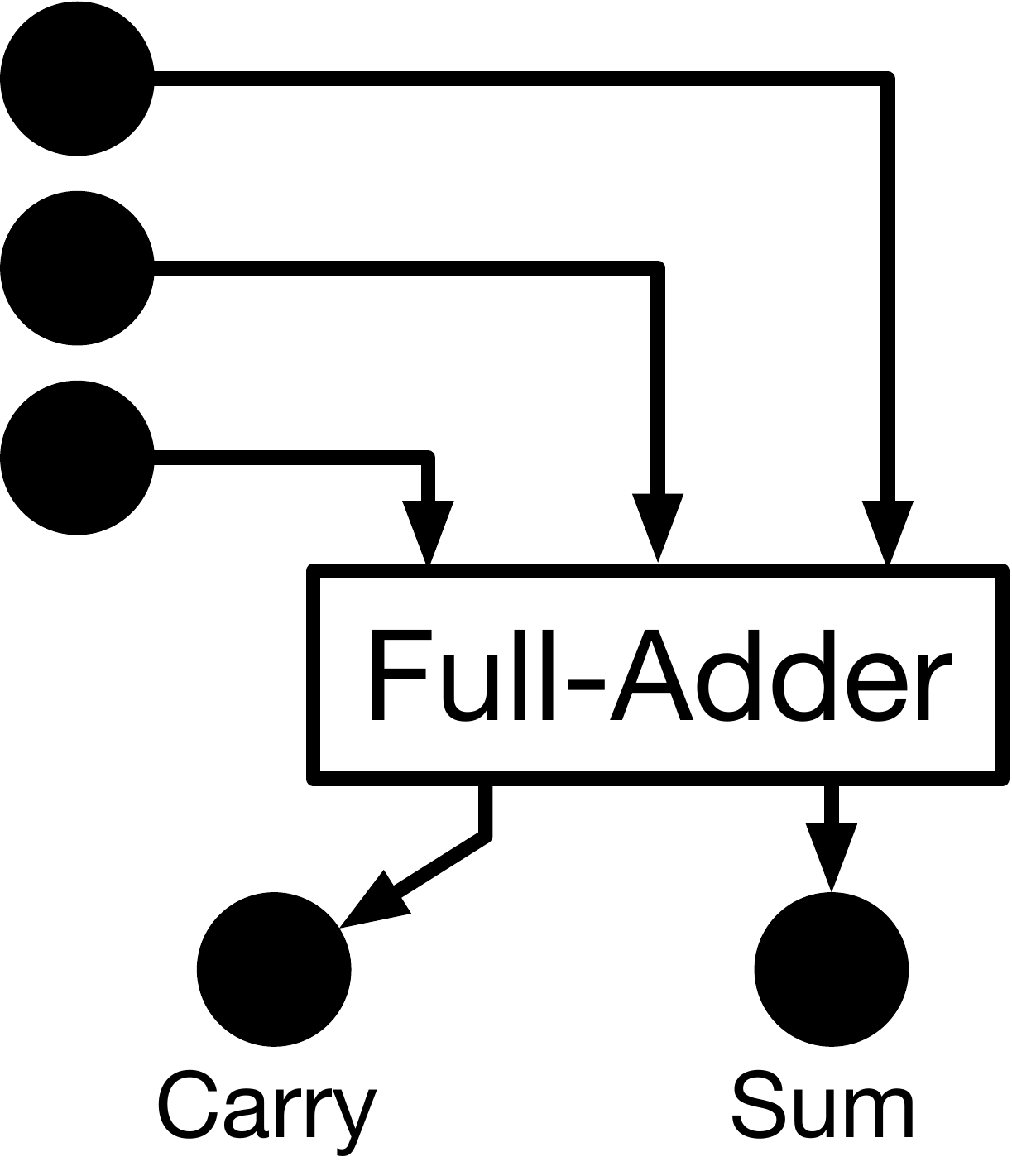}
        \label{dot_example1.fig}
    }\hspace{0.08\linewidth}
    \subfloat[][One stage of 3b carry-save addition of three operands using three FAs in parallel] {
        \centering
        \includegraphics[width=0.40\linewidth]{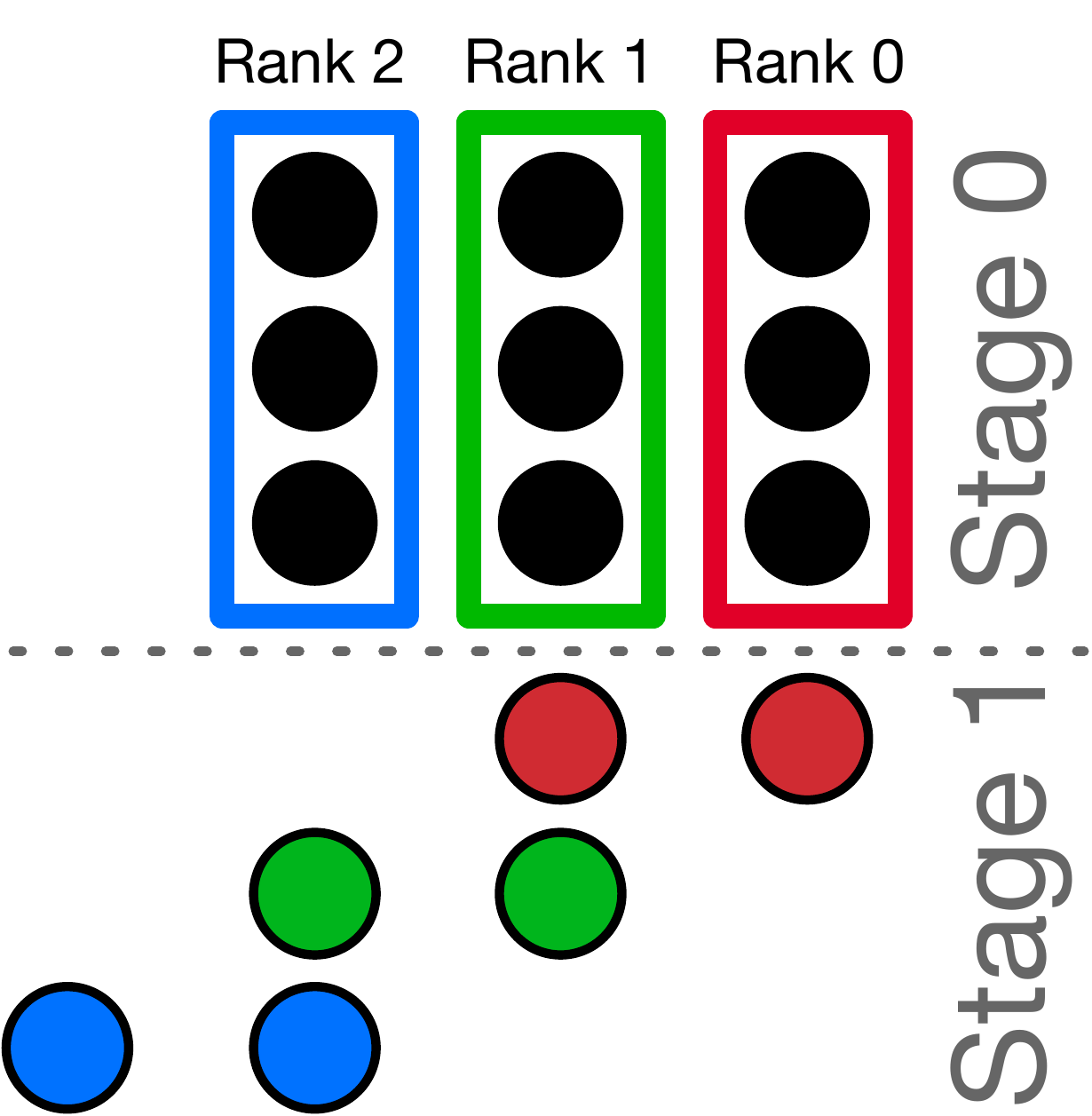}
        \label{dot_example2.fig}
    }
    \vspace{-5pt}
    \caption{($3\text{:}2$) parallel counter, also known as a full-adder.}
    \label{dot_example.fig}
\end{figure}

Parallel counters are digital circuits that simply count the number of asserted bits
in the input, returning this value as a binary output. They can be
specified in ($p\text{:}q$) notation, where $p$ is the number of input bits,
and $q$ is the number of output bits used to express the result in binary
notation. Half-adders (HA) and full-adders (FA) are commonly used
parallel counters, denoted as ($2\text{:}2$) and ($3\text{:}2$) respectively.
Parallel counters can also be expressed in dot
notation~\cite{ercegovac2004digital} as shown for the full-adder in
Figure~\ref{dot_example1.fig}. We use this notation frequently in this
paper to visualize various designs, and use the terms bits and dots
interchangeably. Figure~\ref{dot_example2.fig} shows how FAs can be used in
parallel to implement a single stage of carry-save addition for a 3-bit (3b) 3-operand
addition. Note that each FA takes inputs from a single-column, and hence, all
input bits to a parallel counter have the same $rank$, {\it i.e.} they all have
the same weight.

\subsection{Generalized Parallel Counters}

\begin{figure}[h]
    \centering
    \subfloat[][C6:111] {
        \includegraphics[width=.65in]{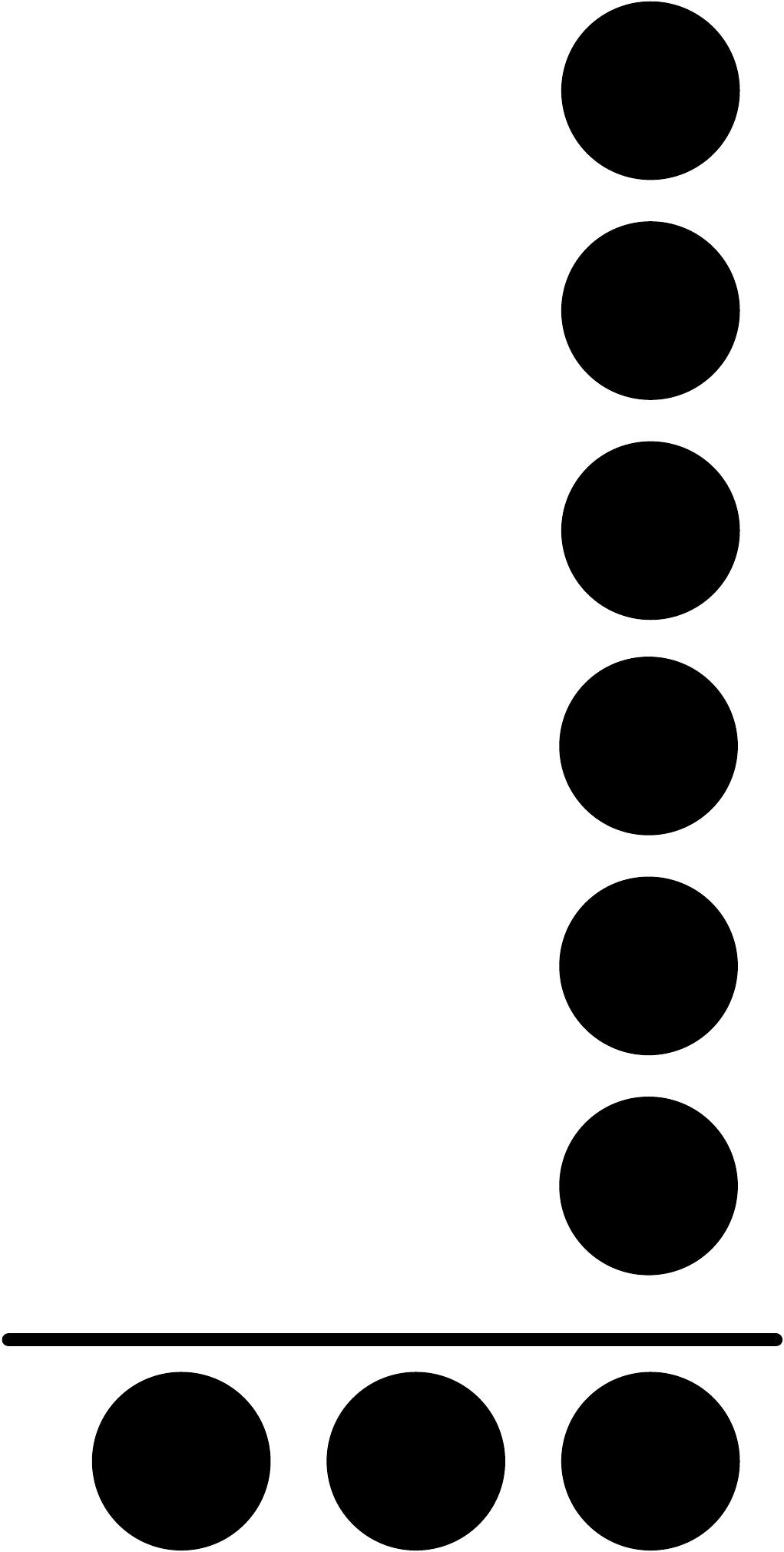}
        \label{c63.fig}
    }\hspace{0.15in}
    \subfloat[][C25:121] {
        \includegraphics[width=.65in]{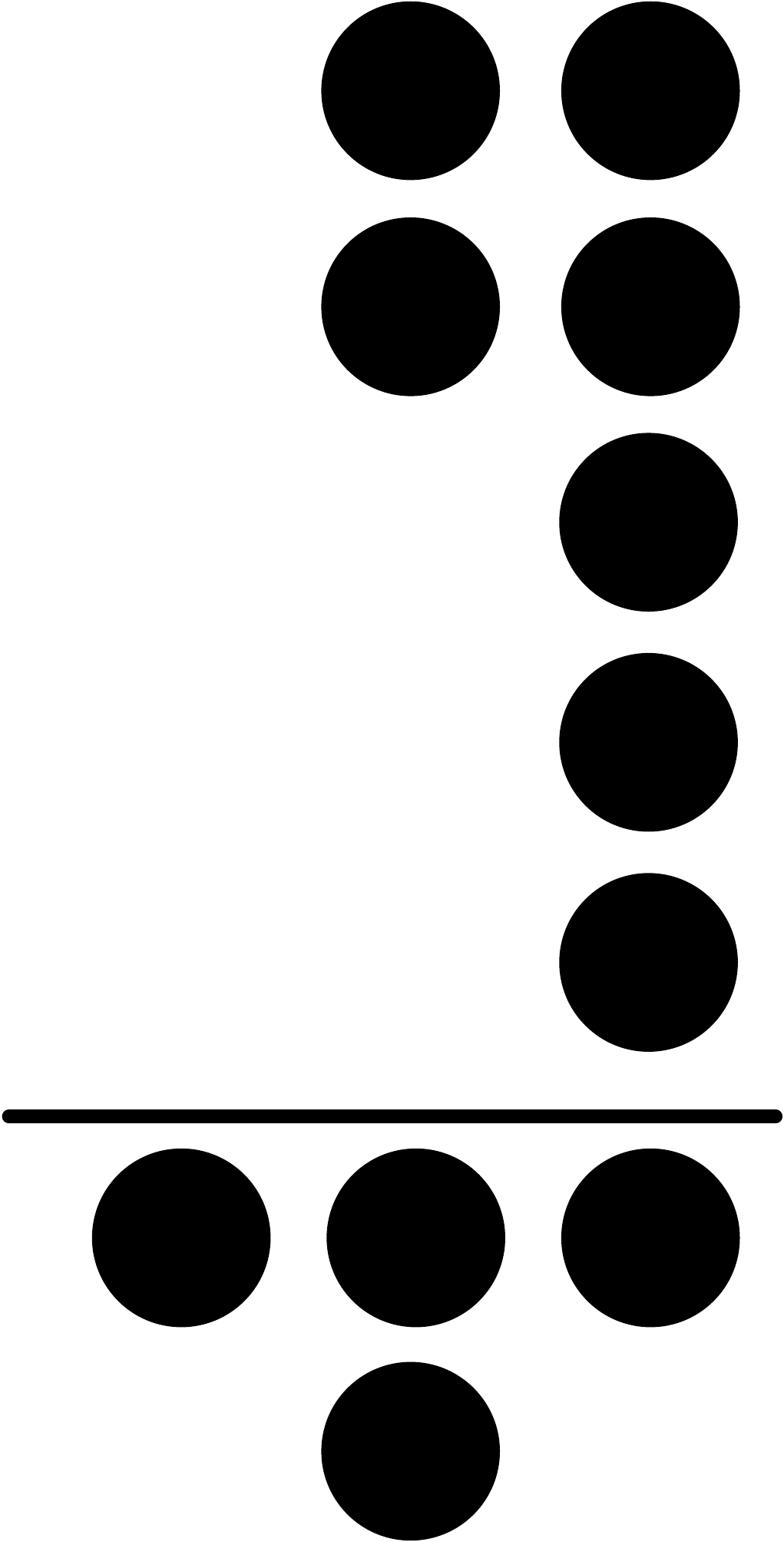}
        \label{c25121.fig}
    }\hspace{0.15in}
    \subfloat[][C1325:11111] {
        \includegraphics[width=1in]{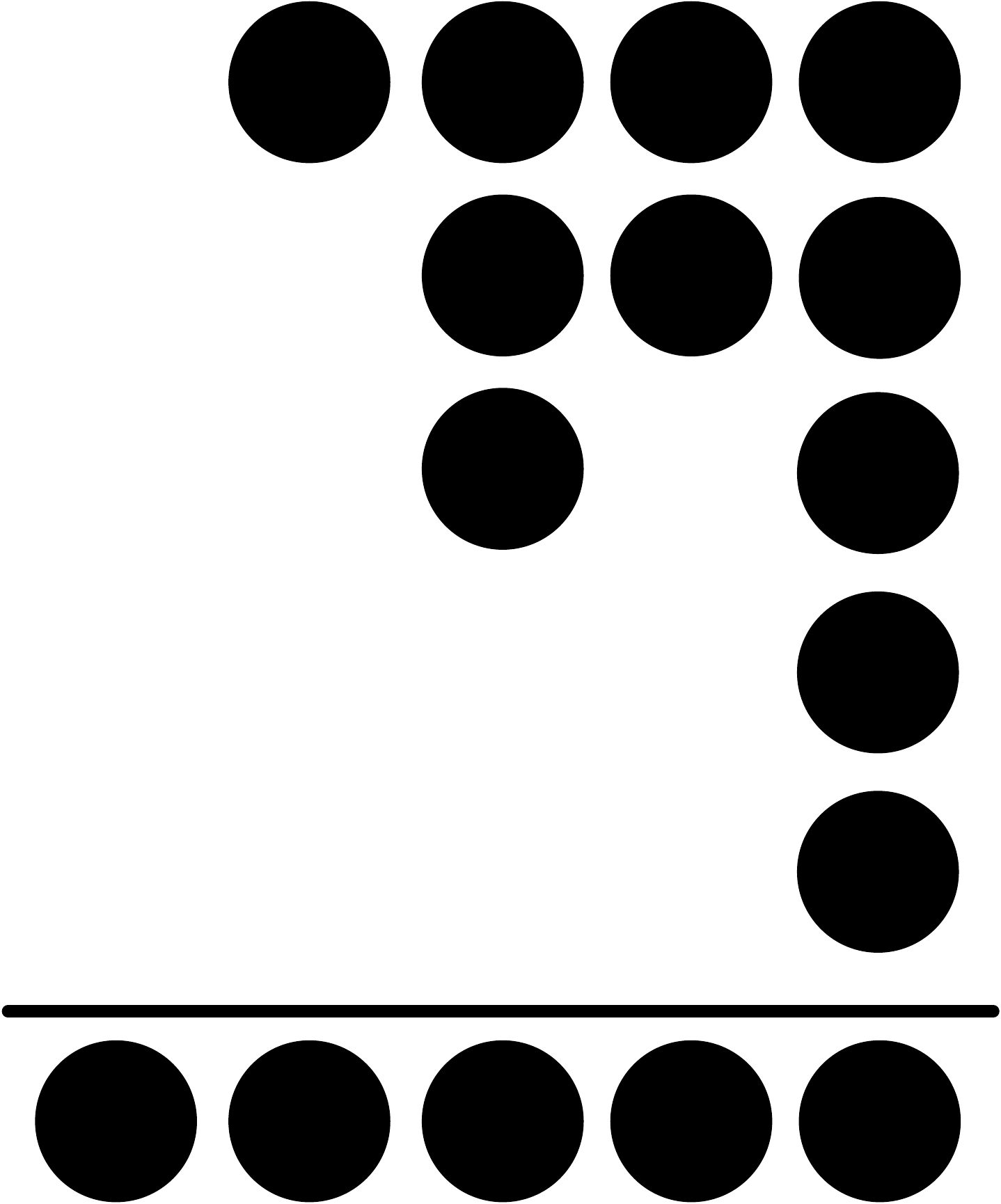}
        \label{c1325.fig}
    }
    \caption{Three popular GPCs found in the literature}
    \label{gpcs.fig}
\end{figure}

Generalized Parallel Counters, or GPCs, were first proposed by Meo~\cite{meo1975arithmetic} and subsequently shown by Parandeh-Afshar et.~al.~\cite{parandeh2008efficient} to map efficiently to FPGAs. Unlike parallel counters, GPCs allow input bits to have different weights, which, in the dot notation, make the GPCs appear as multi-column counters. Figure~\ref{gpcs.fig} shows the dot notation of some previously published GPCs~\cite{DBLP:journals/corr/abs-1806-08095}. Mathematically, GPCs are written as a tuple:
($p_{n-1},\text{...},p_1,p_0\text{:}q_{m-1},\text{...},q_{1},q_{0}$), where $p_i$
is the number of input bits in the $i^{th}$ column, and $q_j$ is the number of output bits in the $j^{th}$ column. FPGA implementations can be classified as lookup table-based GPCs \cite{kumm2018advanced}, or carry-chain-based GPCs \cite{DBLP:conf/fpl/Parandeh-AfsharBI09}. As their names suggest, the ``shape'' of a GPC can have a profound impact on its hardware implementation on FPGAs, and subsequently its performance and efficiency in a compressor tree. Popular metrics to quantify the efficiency of a GPC include \cite{DBLP:journals/trets/Parandeh-AfsharNBI11,DBLP:journals/corr/abs-1806-08095}:
\begin{align*}
    & \text{GPC efficiency, E} = \frac{p-q}{k}\numberthis \label{equ_GPC_E}\\
    & \text{Strength, S} = \frac{p}{q}\numberthis \label{equ_GPC_S}\\
    & \text{Area-Performance Degree, APD} = \frac{(p-q)^2}{k*d} \numberthis \label{equ_GPC_APD}\\
    & \text{Arithmetic slack, A} = 1 - \frac{1+\sum_{i=0}^{m-1}{2^{i} p_{i}}}{1+\sum_{i=0}^{n-1}{2^{i}q_{i}}}\numberthis \label{equ_GPC_A}
\end{align*}
where $p$ and $q$ are the number of input and output bits to/from the GPC respectively, $k$ is the area utilization (in LEs) of the GPC, and $d$ is the critical path delay (in nanoseconds) of the GPC implementation. We tabulate the efficiency of each GPC studied in this work using these metrics later in this paper (Table~\ref{table_Xilinx_new_slice_based_compressor} and Table~\ref{table_intel_GPCs}).

\subsection{Compressors}

\begin{figure}[h]
    \centering
    \subfloat[][Block diagram] {
        \centering
        \includegraphics[width=1.3in]{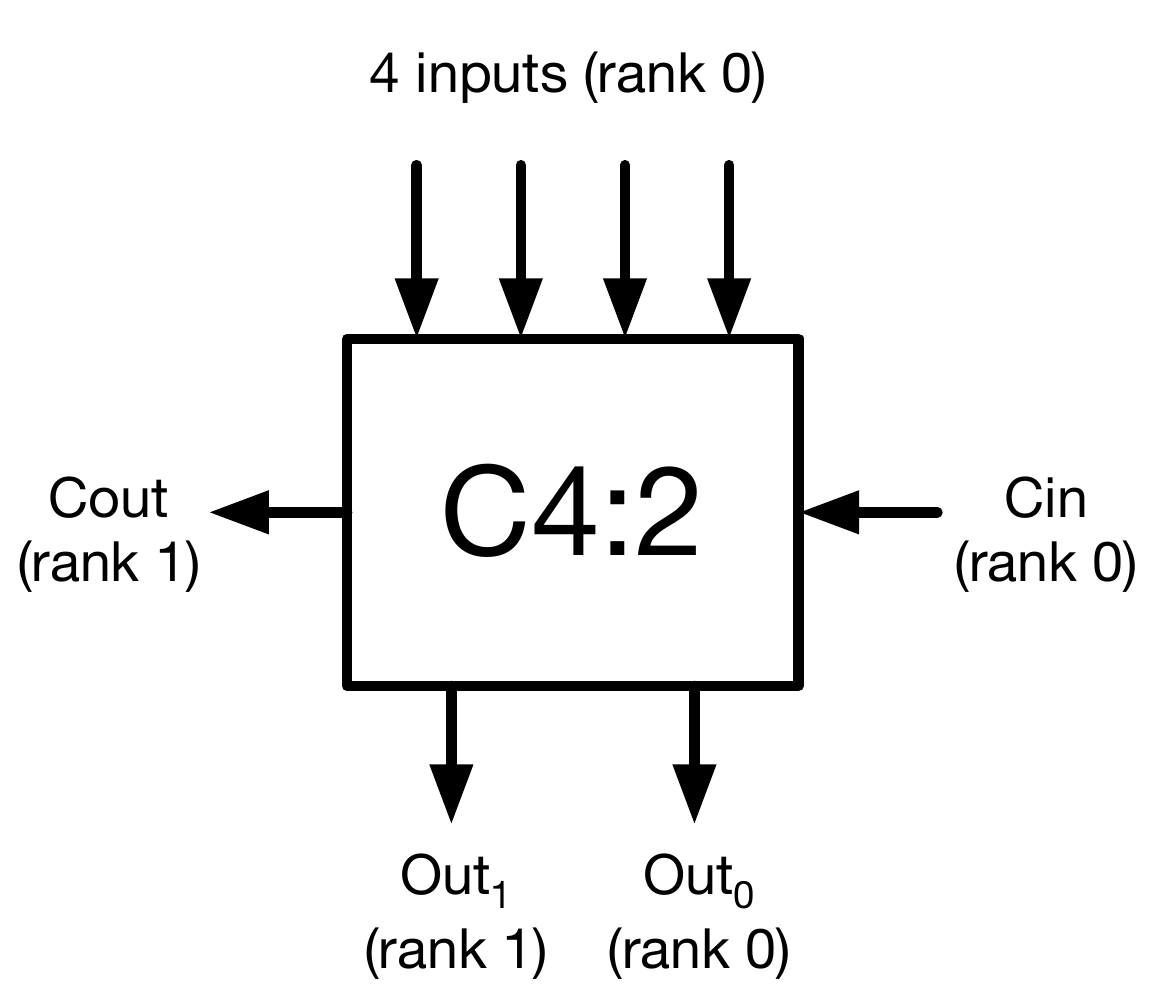}
        \label{compressor42.fig}
    }\hspace{0.3in}
    \subfloat[][One stage of a four-operand 4b-addition using four ($4\text{:}2$) compressors] {
        \centering
        \includegraphics[width=1.3in]{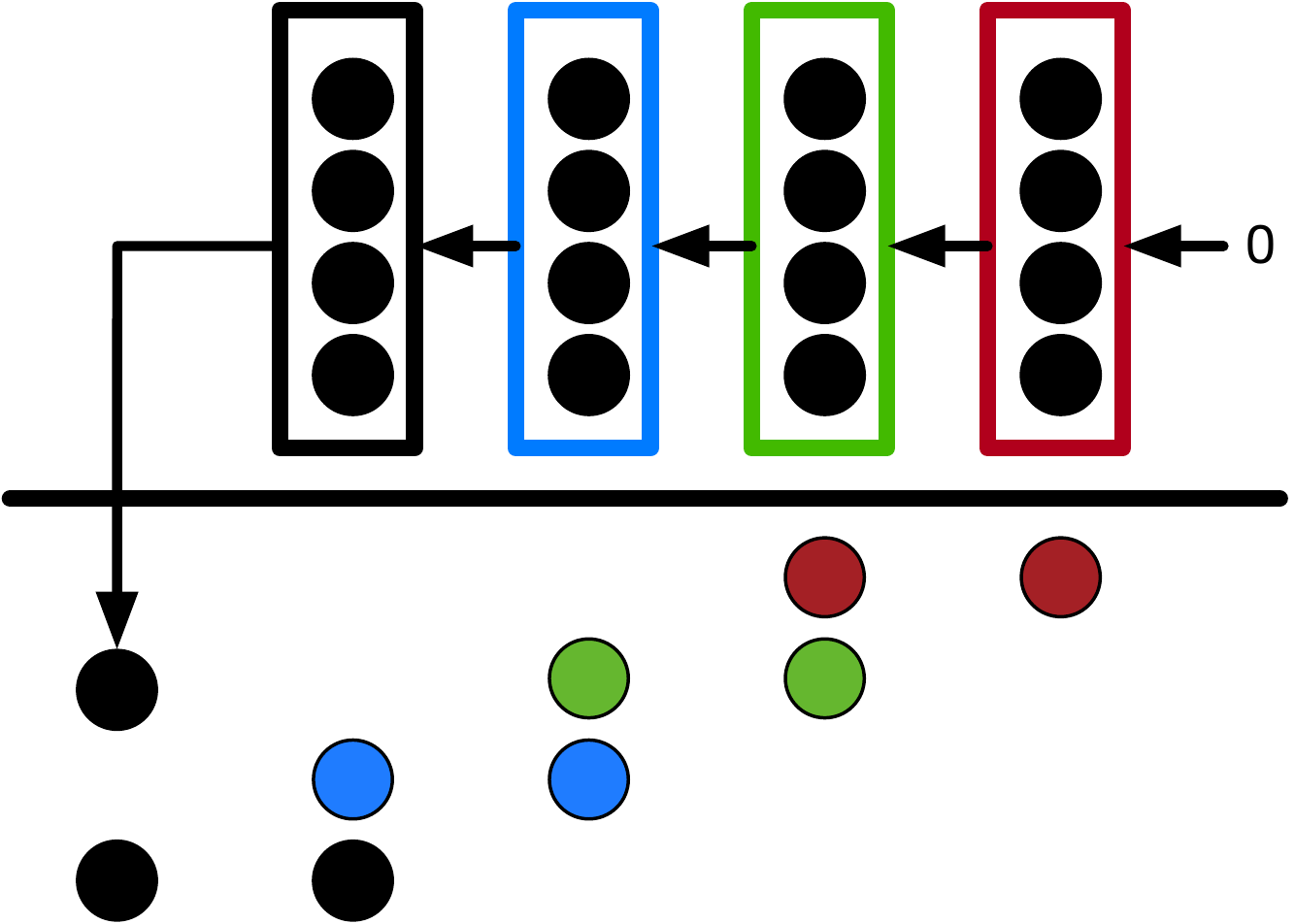}
        \label{compressor_example.fig}
    }
    \vspace{-5pt}
    \caption{Simple ($4\text{:}2$) compressor example.}
    \label{compressors.fig}
\end{figure}

Compressors can be considered parallel counters, with one main difference: they have explicit carry-in (Cin) and carry-out (Cout) bits that can be connected to adjacent compressors in the same stage, as shown in Figure~\ref{compressor_example.fig}. In contrast to carry-propagate adders, the carry-chains between compressors are not cascaded and hence reduce the critical path. Instead they are connected in a carry-save manner. So the overall delay of the circuit scales much better (Figure~\ref{adders.fig}). To the best of our knowledge, the ($4\text{:}2$) compressor (see Figure~\ref{compressor42.fig}) is the only FPGA-friendly~\cite{DBLP:conf/mbmv/KummZ14}) design that targets Xilinx FPGAs, while no efficient compressors exist for Intel devices. Parandeh-Afshar~et~al.~\cite{DBLP:journals/trets/Parandeh-AfsharBI09} addressed this issue by proposing configurable carry-chains as modifications to the Intel Adaptive Logic Module (ALM), supporting 6:2 and/or 7:2 compressors.


For brevity, we describe adders/compressors/GPCs with a simplified notation omitting commas. For example, we describe the GPC (6:1,1,1) as C6:111, the (4:2) compressor as C4:2, or the full adder (3:1,1) as C3:11.

%

\subsection{Adder and Compressor Trees}
\begin{figure}[h]
    \centering
    \subfloat[][Ripple-carry adder.] {
        \centering
        \includegraphics[width=1.4in]{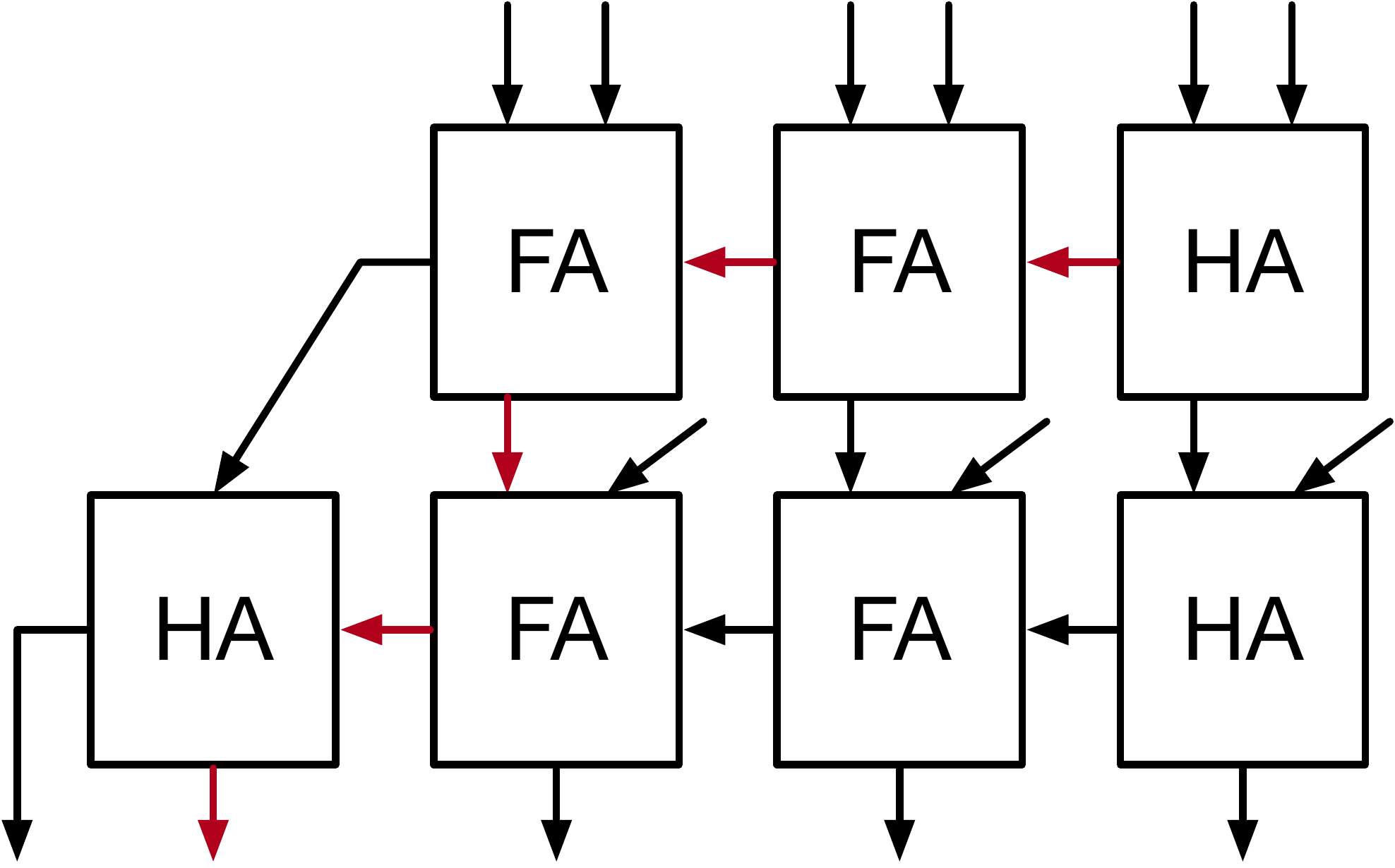}
        \label{rca.fig}
    }
     \hspace{0.1in}
    \subfloat[][Carry-save adder.] {
        \centering
        \includegraphics[width=1.5in]{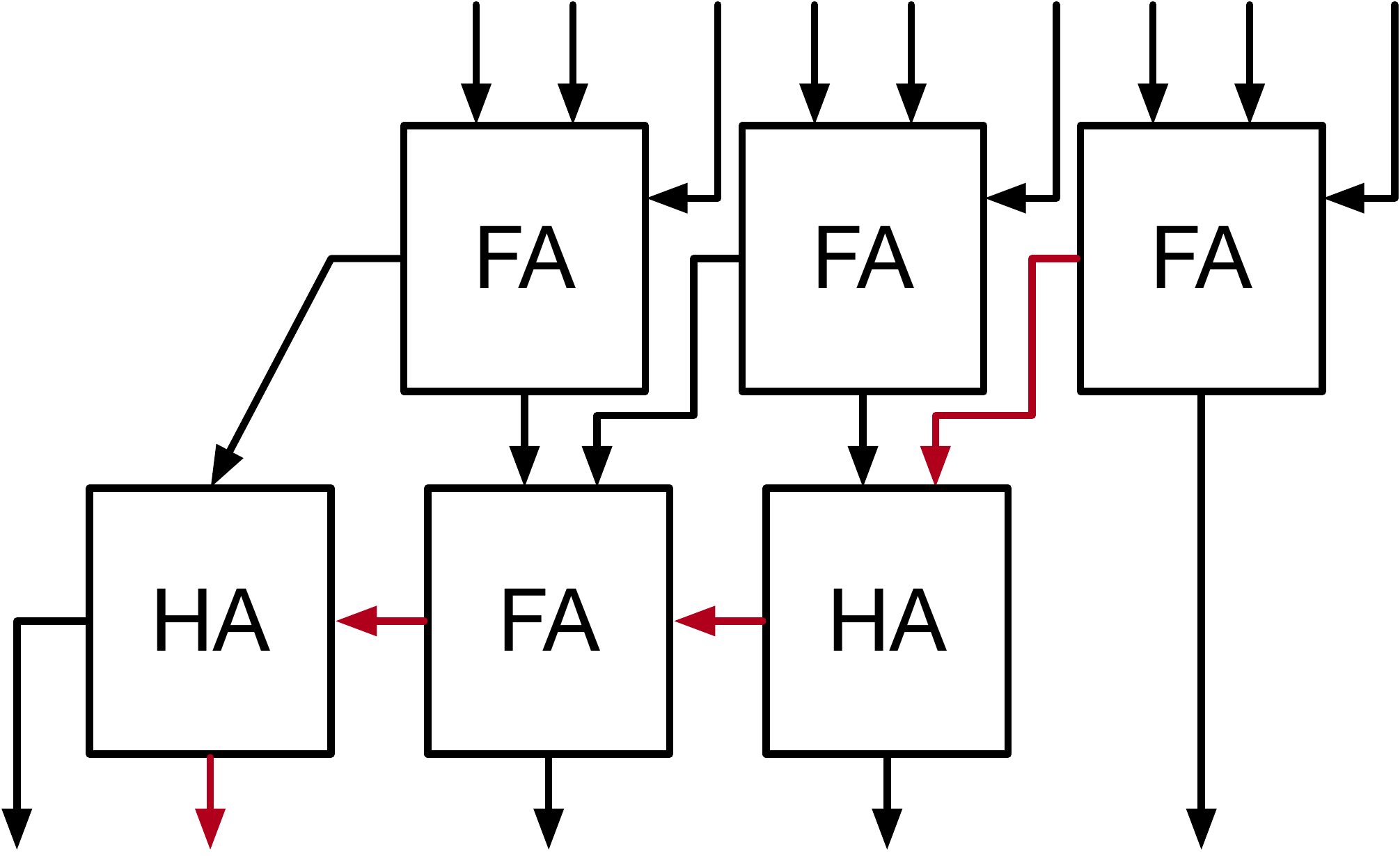}
        \label{csa.fig}
    }
    \caption{Examples of two types of adders.}
    \label{adders.fig}
\end{figure}

For multi-operand addition, we can build adder trees by chaining multiple ripple-carry adders (RCA). Figure~\ref{rca.fig} shows addition of 3$\times$3b operands. The carry-out from each FA/HA propagates to the next FA, which results in a long critical path along the carry-chain (shown in red). While the RCA has a small area footprint, this long delay is undesirable and can limit performance, especially for operands with large bitwidth.

The carry-save adder (CSA)~\cite{earle65} addresses this issue by treating the full-adder as a C3:11 compressor and breaking the carry-chain as shown in Figure~\ref{csa.fig}. By avoiding the carry chain, the delay is largely determined by the depth of the tree. 
However, the final stage must be reduced to the final answer using an RCA. Nevertheless, the CSA adder reduces the overall delay of the addition. For the example in Figure~\ref{adders.fig}, the critical path delay has one less full-adder-delay.

\begin{figure*}[ht!]
    \subfloat[][Modifications to Xilinx UltraScale+ LE. A slice is composed of four LEs.] {
        \centering
        \includegraphics[width=3.8in]{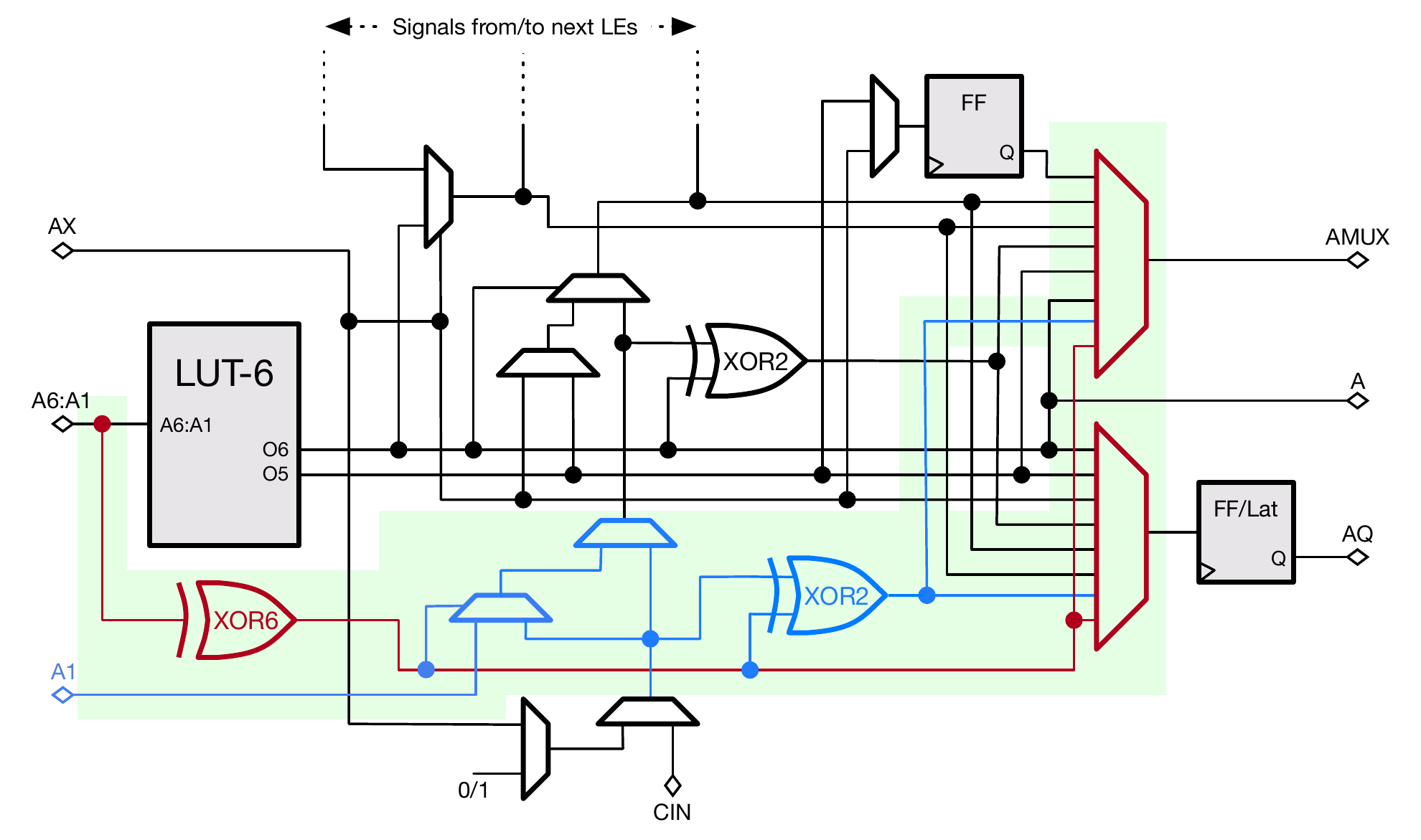}
        \label{xilinx_slice_mod.fig}
    }\hspace{0.05in}
    \subfloat[][Modifications to the Intel Stratix-10 ALM. Each ALM has 8-inputs and a fracturable LUT6~\cite{ug_s10_lab}.] {
        \centering
        \includegraphics[width=3in]{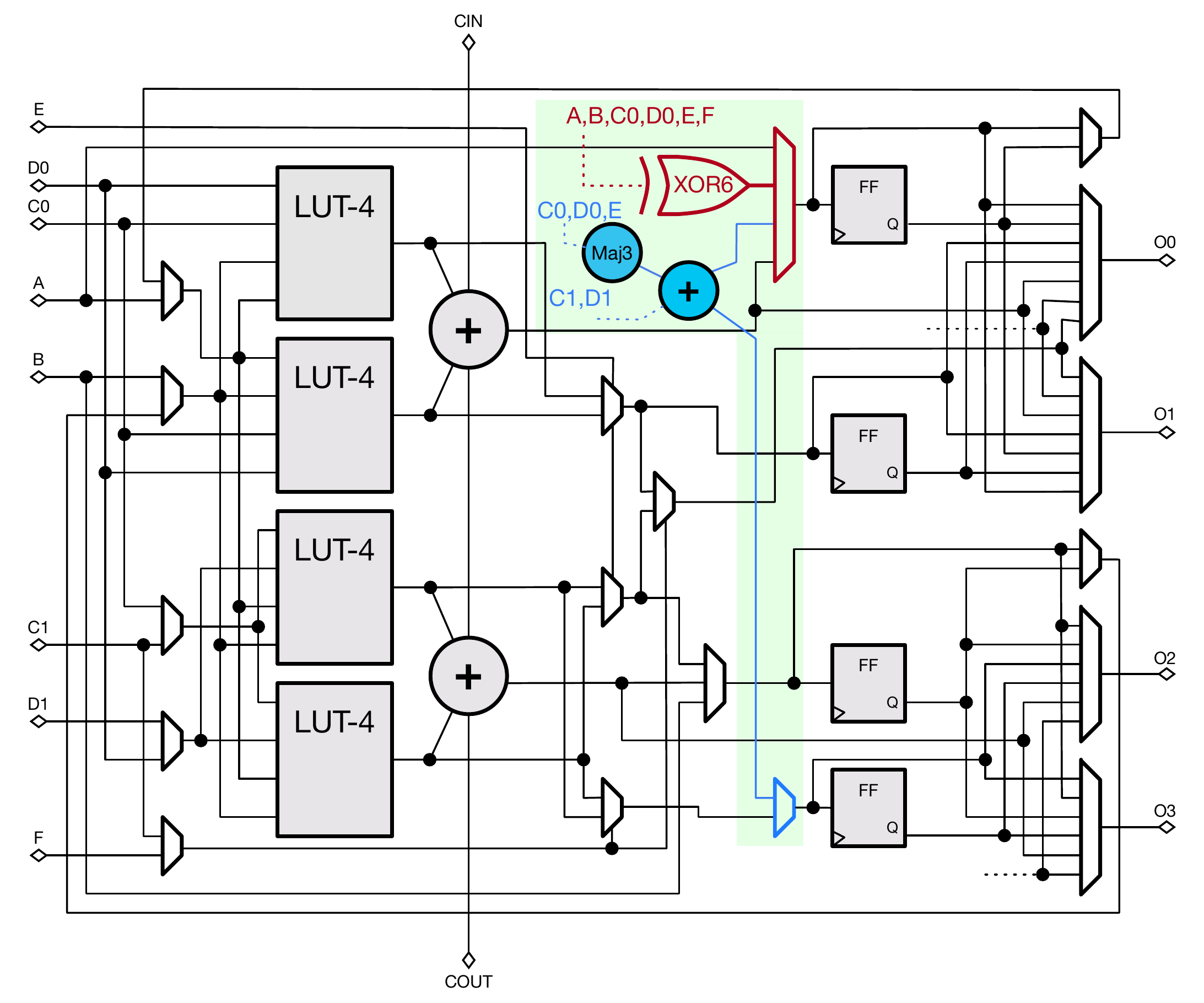}
        \label{intel_alm_mod.fig}
    }
    \caption{Basic logic element (LE) for Xilinx and Intel FPGA architectures. LUXOR modifications are highlighted in red, while vendor-specific LUXOR+ modifications are colored blue. Some signals are omitted for simplicity.}
    \label{arch_top.fig}
\end{figure*}
   
This idea of breaking the carry-chain dependency up till the final RCA stage is the basis behind compressor trees. A compressor tree is simply a circuit that takes in a set of binary values (or dots) that represent multiple operands, and outputs the result as a sum and carry. Stage 0 in Figure~\ref{dot_example2.fig} is a compressor tree that produces sum and carry bits as inputs into Stage 1, which are then evaluated by an RCA to produce the final result (see HA$\rightarrow$FA$\rightarrow$HA row in Figure~\ref{csa.fig}, which is the RCA stage). Compressor trees can be built using GPCs, compressors, or both, and efficient compressor tree design is an active area of research with large bodies of existing literature ~\cite{DBLP:conf/mbmv/KummZ14,DBLP:journals/trets/Parandeh-AfsharNBI11,parandeh2009improving,DBLP:conf/fpl/Parandeh-AfsharBI09, DBLP:conf/date/VermaI07,DBLP:journals/trets/Parandeh-AfsharBI09, DBLP:journals/corr/abs-1806-08095,DBLP:conf/fpl/KummZ14}. 

The reader is encouraged to read~\cite{DBLP:journals/trets/Parandeh-AfsharBI09} for a more detailed background on parallel counters, GPCs, compressors, and different methods of compressor tree implementations.



\subsection{Xilinx FPGA Logic Elements}

The Xilinx configurable logic block (CLB)~\cite{ug574} is composed of two slices, which are the basic unit of the FPGA's soft-fabric. 
Each slice is composed of four 6-input LEs, including 6-input LUT and additional circuitry such as registers and multiplexers, which give the slice its expressiveness. Figure~\ref{xilinx_slice_mod.fig} (in black) shows a quarter of the slice architecture (an 6-input LE and the corresponding circuits) found in the modern Xilinx UltraScale+ FPGAs. Another notable feature of the slice is the presence of a fast carry-chain between the LEs, which is often used to implement arithmetic circuits such as RCAs. The architectural modifications proposed in this work are at the \emph{quarter slice} abstraction.

\subsection{Intel FPGA Logic Elements}

The main logic element in Intel FPGAs is the adaptive logic module (ALM)~\cite{ug_s10_lab}. Figure~\ref{intel_alm_mod.fig} (in black) shows the ALM architecture of a modern Stratix-10 device. Each ALM is composed of a fracturable 6-input LUT, while primitives such as full-adders and multiplexers help to support higher-order boolean functions. Ten ALMs on Intel FPGAs are grouped to form a logic array block (LAB), which augments the ALMs with more primitives such as HyperFlex registers, local interconnect, and configurable carry-chains~\cite{ug_s10_lab}. Our proposed modifications in this paper are at the \emph{ALM} abstraction.

\subsection{Related Work}

Parallel digital arithmetic circuits have been explored since the
1960s~\cite{wallace1964suggestion,dadda1965some,swartzlander1973parallel}, but FPGA-based compressor trees were only popularized in the past two decades, primarily from work by Parandeh-Afshar~et.~al.~\cite{parandeh2008efficient,DBLP:journals/trets/Parandeh-AfsharBI09} and Kumm~et.~al.~\cite{DBLP:conf/mbmv/KummZ14,DBLP:conf/fpl/KummZ14}. In~\cite{DBLP:journals/trets/Parandeh-AfsharBI09}, the authors proposed architectural changes to the Intel ALM carry-chains such that large compressors like ($6\text{:}2$) and ($7\text{:}2$) can be efficiently mapped to single ALMs. Although their proposed compressor is very
efficient, for modern applications such as BNN popcounting~\cite{liang2018fp}, these compressors
would be significantly underutilized. Similarly, Kim et.
al.~\cite{DBLP:conf/fpt/KimLA18} and Boutros et. al.~\cite{boutros2019math} propose
changes to the FPGA architecture, by adding sum-chain and extra carry chains respectively, specifically for modern deep neural network
applications, which do not necessarily benefit general-purpose compressor
trees. Our proposed changes are motivated by insight into modern
GPC-based compressor tree designs, and benefit a larger suite of old and new benchmarks.



\section{FPGA Logic Cell Enhancements}
\label{se:luxor}

In this section, we describe in detail the proposed hardware architecture modifications that further improve the performance of GPCs on FPGAs. We focus our efforts on improving the design of the logic cell of FPGAs from the two major FPGA vendors, Intel and Xilinx. Our modifications are organized into two tiers: (1) A vendor-agnostic change to both Intel and Xilinx FPGA logic cells, and (2), a vendor-specific modification on top that further optimizes performance. We refer to these logic cell design tiers as LUXOR, and LUXOR+ respectively. Both LUXOR and LUXOR+ are backward-compatible and retains pin-interchangeability, {\it i.e.} any existing design maps equally well to these new architectures.

\subsection{LUXOR}
\label{se:bnn}
\begin{figure}[t]
    \centering
    \includegraphics[width=\linewidth]{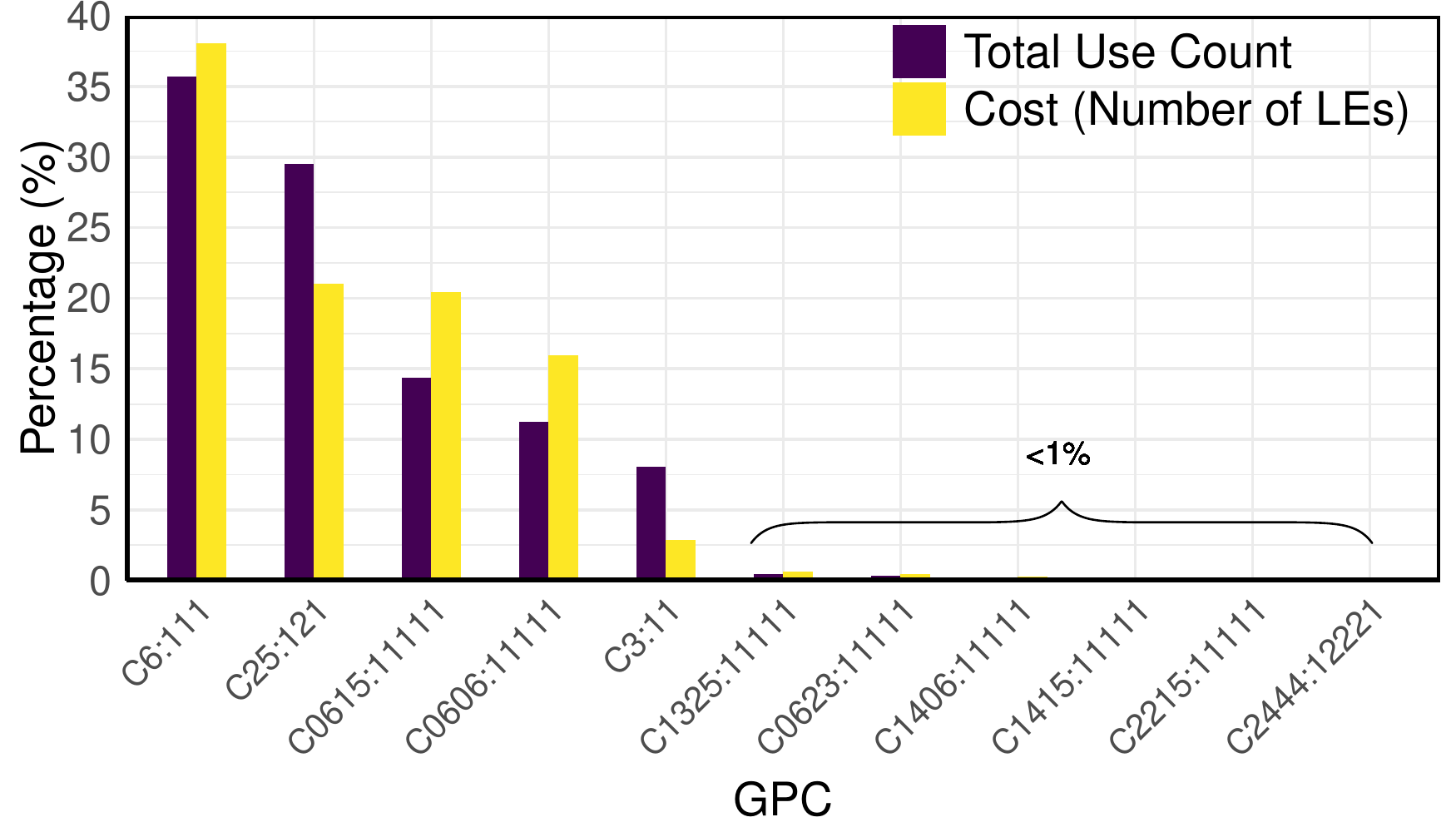}
    \vspace{-20pt}
    \caption{Total percentage count and cost of each GPC/compressor found in optimal solutions of compressor trees across 50$+$ Micro-Benchmarks from a variety of fields. The GPC/compressor list is according to~\cite{DBLP:journals/corr/abs-1806-08095}}
    \label{dist_gpcs.fig}
\end{figure}

Our first proposed modification is to add a 6-input XOR gate (XOR6) to both Intel and Xilinx FPGA cells. The XOR6 is parallel to the LUT and re-uses its inputs and output path as shown in Figure~\ref{arch_top.fig}.
This modification is motivated by the observation that the C6:111 GPC is dominant in modern FPGA-based compressor tree designs. To quantify that claim, we analyzed optimal solutions of compressor trees from a set of 50$+$ micro-benchmarks that are commonly found in various domains ({\it e.g.} popcounting, multi-operand addition, FIR filters, etc) using efficient GPCs and compressors for Xilinx architecture from reference~\cite{DBLP:journals/corr/abs-1806-08095}. Figure~\ref{dist_gpcs.fig} shows a histogram of the percentage count and cost (in LEs) for all GPCs across all solutions. Due to its compression efficiency, C6:111 is used more than a third of the time, and as a result, most of the hardware is dedicated towards its implementation. In modern FPGAs, the C6:111 maps to 3 LUTs, but by providing an explicit XOR6 datapath inside each logic cell, we can bring that cost down to 2 LEs. This is done by mapping the first output bit to the XOR6 rather than using a separate LE. Hence, LUXOR can deliver a resource utilization reduction for the most commonly-used GPC of up to 33\%.

Another very useful feature of the LUXOR design is its applicability to binarized neural networks (BNNs). In BNNs, the core computational workload is generated by the convolution layers, which are reduced via a XnorPopcount~\cite{rastegari2016xnor} operation for the binary case. Consider the XnorPopcount operation between three binary activations ($x_0$, $x_1$, and $x_2$) and their corresponding binary weights ($w_0$, $w_1$, and $w_2$). The required computation is:
\begin{align*}
    & \text{Sum} = (w_0 \overline\oplus x_0) \oplus (w_1 \overline\oplus x_1) \oplus (w_2 \overline\oplus x_2)\\
    & \text{Carry, C} = (w_0 \overline\oplus x_0) \cdot (w_1 \overline\oplus x_1) + (w_2 \overline\oplus x_2)[(w_0 \overline\oplus x_0) \oplus (w_1 \overline\oplus x_1)]
\end{align*}
where $\overline\oplus$ and $\oplus$ represent the XNOR and XOR operations respectively.

\begin{figure}[ht!]
    \centering
    \subfloat[] {
        \includegraphics[width=1.35in]{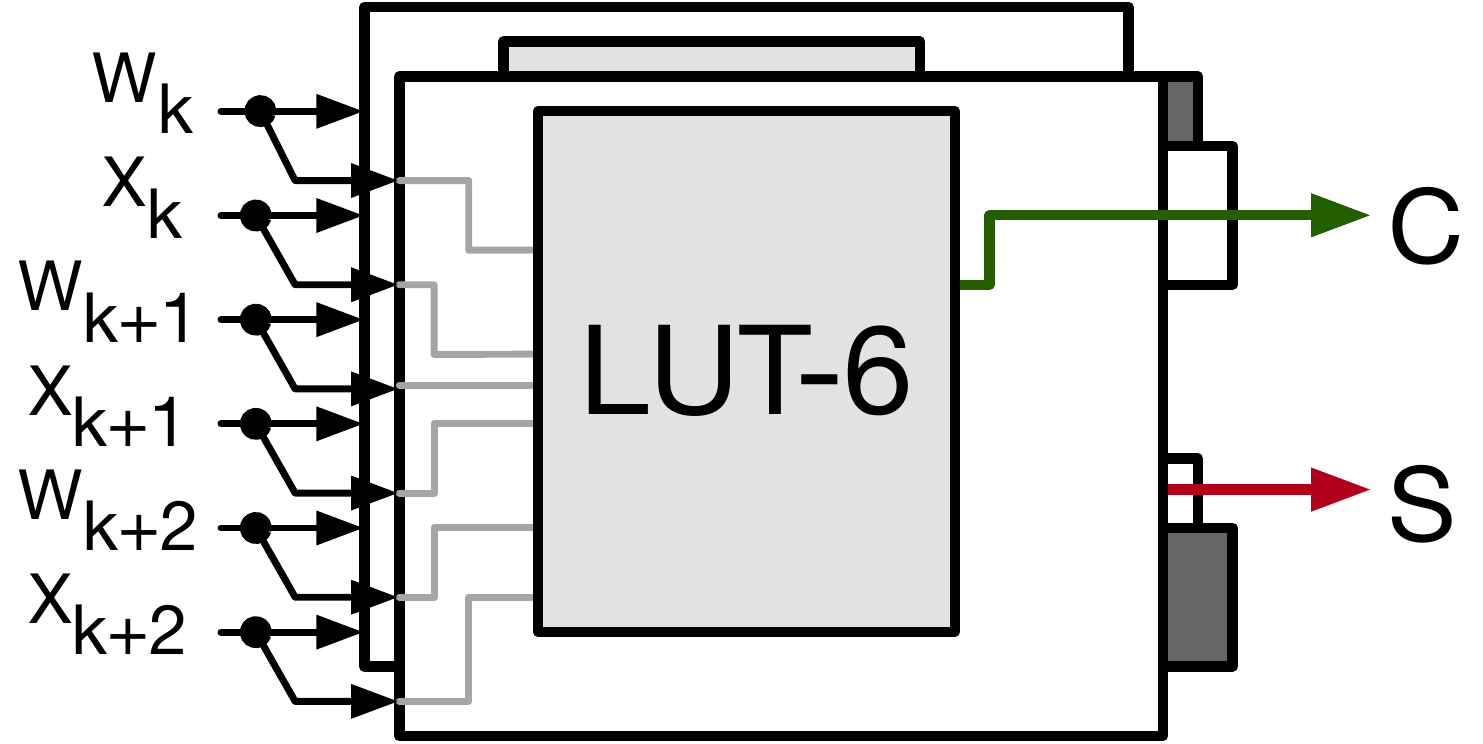}
        \label{bnn_2luts.fig}
    }\hspace{0.01in}
    \subfloat[] {
        \includegraphics[width=1.45in]{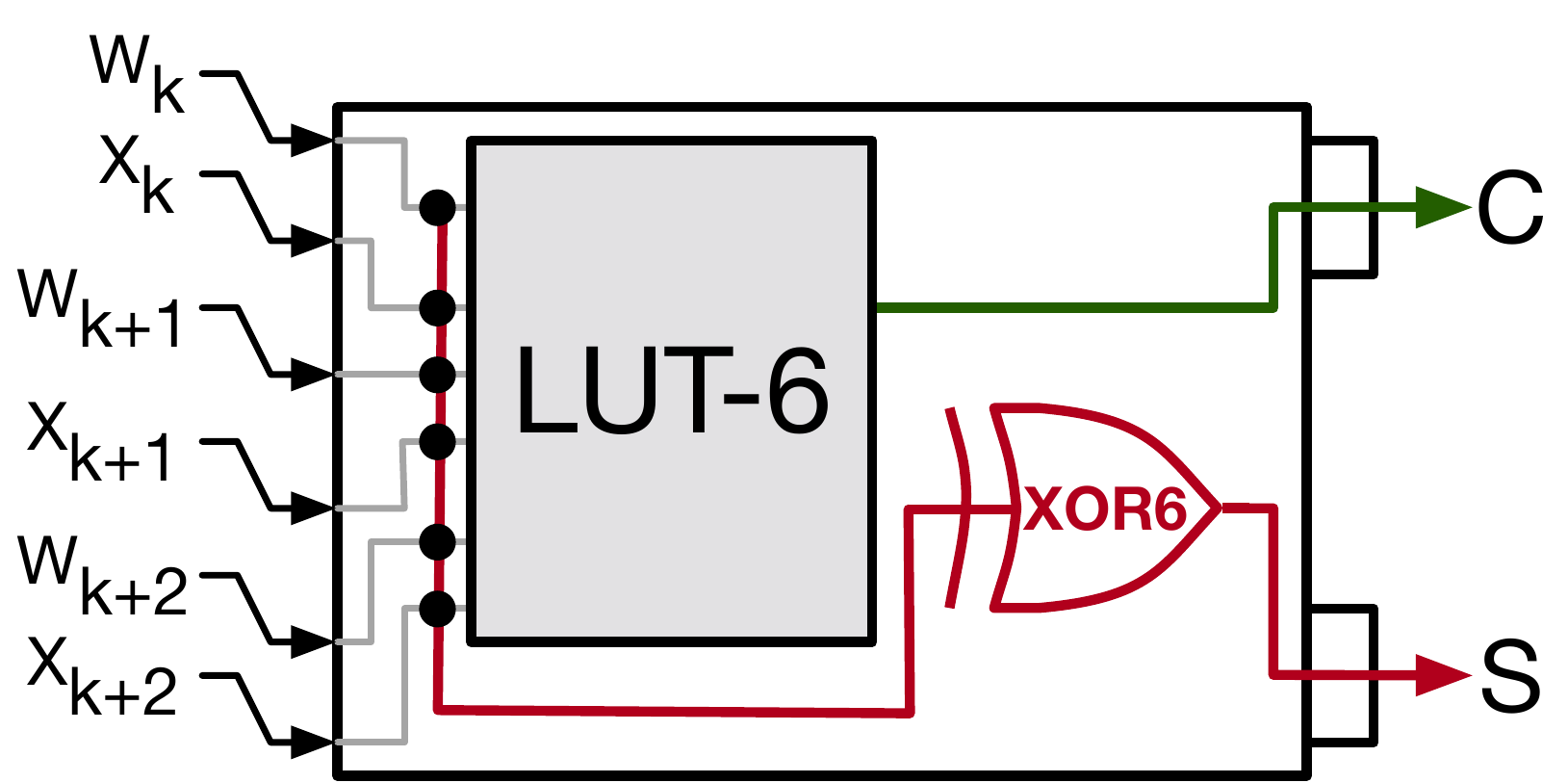}
        \label{bnn_luxor.fig}
    }\hspace{0.01in}
    \subfloat[] {
        \includegraphics[width=0.25in]{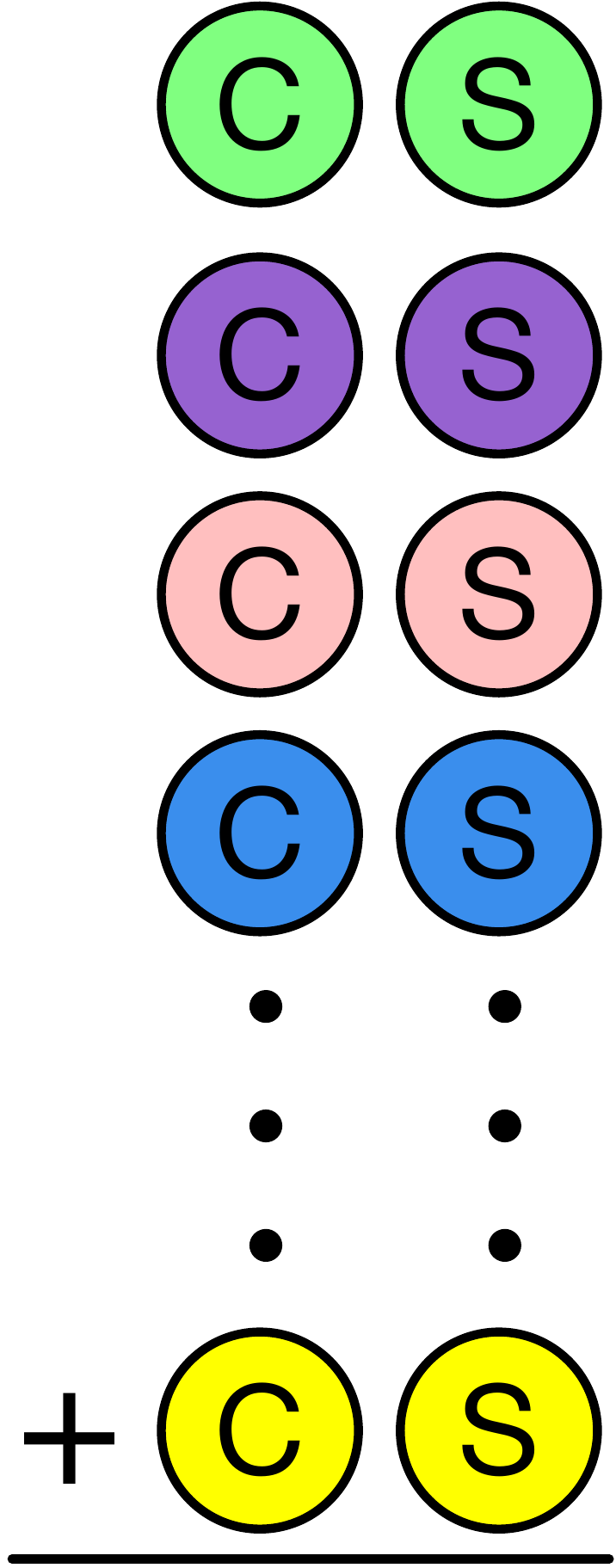}
        \label{bnn_2col.fig}
    }
    \caption{BNN implementation on Xilinx FPGAs: primary multiply and compressors of (a) XnorPopcount with 2 LEs, (b) XnorPopcount with 1 LUXOR LE. (c) Final two-column popcount to accumulate the partial sums (S), and carries (C)}
    \label{bnn_idea.fig}
\end{figure}

This XnorPopcount operation gets mapped to 2 LEs on modern FPGAs, as shown in Figure~\ref{bnn_2luts.fig} -- one LE to compute the sum bit, and the other to compute the carry bit. With LUXOR, however, this computation can be mapped to just a single logic element via a Boolean transformation, where the Sum bit (S) can now be expressed as:
\begin{align*}
    & \text{Sum} = (\overline{w_0} \oplus x_0) \oplus (\overline{w_1} \oplus x_1) \oplus (\overline{w_2} \oplus x_2)
\end{align*}
which is essentially a XOR6 function where the complement of the weights are used. The LUT-6 implements the carry logic in this case, and both outputs from a single Xilinx slice can now be used to compute the partial products of the binarized convolution layer (see Figure~\ref{bnn_luxor.fig}). Finally, to compute the output activations of the convolution layer, all the partial sums have to be summed, which can be visualized as a tall two-column many-operand instruction of carry and sum bits, as shown in Figure~\ref{bnn_2col.fig}. This can be efficiently reduced using a compressor tree, which is also improved by our proposed LUXOR modifications.

\vspace{20pt}
\subsection{LUXOR+}

\subsubsection{LUXOR+ for Xilinx FPGAs (X-LUXOR+)}
\label{sssec: Xlux+}

Reference~\cite{DBLP:journals/corr/abs-1806-08095} proposes the {\em atoms} (--06--, --14--, --22--), as primitives to construct slice-based GPCs. Atoms are 2-column-input GPCs which mapped well to half of a slice (2 LEs) and can be connected via fast in-slice carry-chains to form wider GPCs, called couple. Note, the first atom in a couple can also accept one extra input in the first rank, except --06-- for structural reasons. For instance --06-- and --22-- atoms builds two couples as C0623:11111 and C2206:11111. All combinations of these three atoms as well as C1325:11111 (which is also a slice-based GPC but not decomposable) are listed in the baseline section of Table~\ref{table_Xilinx_new_slice_based_compressor}. 

The blue datapath in Figure~\ref{xilinx_slice_mod.fig} highlights the proposed modification to the Xilinx FPGA slice. It involves modification of the carry chain datapath, introducing additional logic to allow the output from the XOR6 gate to be propagated into the carry-chain. This allows us to improve the implementation of slice-based GPCs. This enables us to map atom --06-- to a quarter slice and consequently offers new set of slice-based GPCs such as C06060606:111111111, which can be mapped to just a single slice. This particular GPC has a very high compression efficiency of 3.75, which is more than any other existing GPCs in the literature. The X-LUXOR+ portion of Table~\ref{table_Xilinx_new_slice_based_compressor} summarizes the characteristics of the new GPCs for Xilinx FPGAs.


\begin{table}[thb]
\renewcommand{\arraystretch}{1.0}
\setlength{\tabcolsep}{4pt}
    \caption{Slice-based GPCs for Xilinx FPGAs. N.B. X-LUXOR+ area overhead is not considered in computing $\bm{E},\bm{S},\bm{A}$.}
\label{table_Xilinx_new_slice_based_compressor}
\vspace{-7pt}
\centering
\begin{tabular}{|c|c|c|c|c|c|c|c|}
    \hline
    \multicolumn{2}{|c|}{\bf GPCs} & {$\bm{p}$} & {$\bm{q}$} & {\bf LUTs} & {$\bm{E}$} & {$\bm{S}$} & {$\bm{A}$}\\
    \hline
    \hline
    \multirow{10}{*}{\rotatebox{90}{\bf Baseline~\cite{DBLP:journals/corr/abs-1806-08095,DBLP:conf/mbmv/KummZ14}}} & {C0606:11111} & {12} & {5} & {4} & {1.75} & {2.40} & {0.031}\\
    \cline{2-8}
    {} & {C1415:11111} & {11} & {5} & {4} & {1.50} & {2.20} & {0.000}\\
    \cline{2-8}
    {} & {C2215:11111} & {10} & {5} & {4} & {1.25} & {2.00} & {0.000}\\
    \cline{2-8}
    {} & {C0615:11111} & {12} & {5} & {4} & {1.75} & {2.40} & {0.000}\\
    \cline{2-8}
    {} & {C1423:11111} & {10} & {5} & {4} & {1.25} & {2.00} & {0.000}\\
    \cline{2-8}
    {} & {C2223:11111} & {9} & {5} & {4} & {1.00} & {1.80} & {0.000}\\
    \cline{2-8}
    {} & {C0623:11111} & {11} & {5} & {4} & {1.50} & {2.20} & {0.000}\\
    \cline{2-8}
    {} & {C1406:11111} & {11} & {5} & {4} & {1.50} & {2.20} & {0.031}\\
    \cline{2-8}
    {} & {C2206:11111} & {10} & {5} & {4} & {1.25} & {2.00} & {0.031}\\
    \cline{2-8}
    {} & {C1325:11111} & {11} & {5} & {4} & {1.50} & {2.20} & {0.063}\\
    \hline
    \hline
    \multirow{9}{*}{\rotatebox{90}{\bf X-LUXOR$+$}} & {C06060606:111111111} & {24} & {9} & {4} & {\bf 3.75} & {\bf 2.67} & {0.002}\\
    \cline{2-8}
    {} & {C140606:1111111} & {17} & {7} & {4} & {2.50} & {2.43} & {0.008}\\
    \cline{2-8}
    {} & {C220606:1111111} & {16} & {7} & {4} & {2.25} & {2.29} & {0.008}\\
    \cline{2-8}
    {} & {C060606:1111111} & {18} & {7} & {4} & {2.75} & {2.57} & {0.008}\\
    \cline{2-8}
    {} & {C060615:1111111} & {18} & {7} & {4} & {2.75} & {2.57} & {0.000}\\
    \cline{2-8}
    {} & {C060623:1111111} & {17} & {7} & {4} & {2.50} & {2.43} & {0.000}\\
    \cline{2-8}
    {} & {C061406:1111111} & {17} & {7} & {4} & {2.50} & {2.43} & {0.008}\\
    \cline{2-8}
    {} & {C062206:1111111} & {16} & {7} & {4} & {2.25} & {2.29} & {0.008}\\
    \hline
\end{tabular}
\end{table}

\begin{figure*}[ht!]
    \centering
    \includegraphics[width=0.80\linewidth]{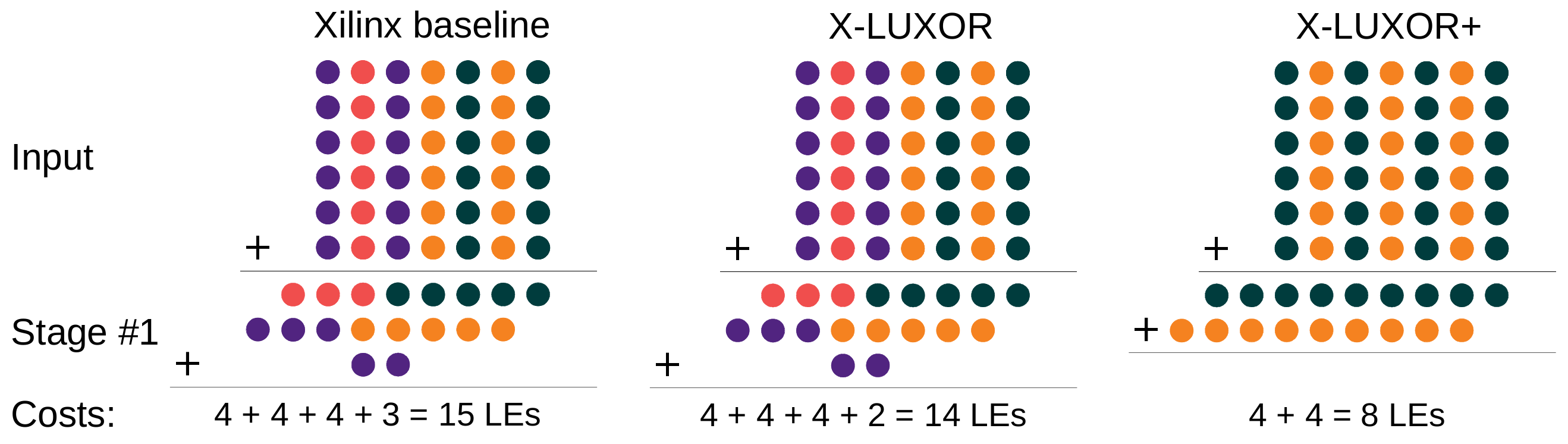}
    \caption{Example compressor tree for a 6-operand 7-bit addition using Xilinx baseline, X-LUXOR, and X-LUXOR+ 
    }
    \label{Dot_new.fig}
\end{figure*}

We provide a simple illustration of the impact of our X-LUXOR and X-LUXOR+ optimizations in Figure~\ref{Dot_new.fig}. The penultimate (red) column can be implemented with a C6:111 compressor, requiring 2 LEs (instead of 3 in the unmodified case) in X-LUXOR. X-LUXOR+ is able to use the C06060606:111111111 GPC, which further reduces resource usage. In general, X-LUXOR has the greatest impact on tall-skinny compressor trees, which require significant use of C6:111, and hence has greater gains for wide compressor trees.

%
%
%

\subsubsection{LUXOR+ for Intel FPGAs (I-LUXOR+)}
\label{se:majfa}

\begin{figure}[ht!]
    \centering
    \includegraphics[width=\linewidth]{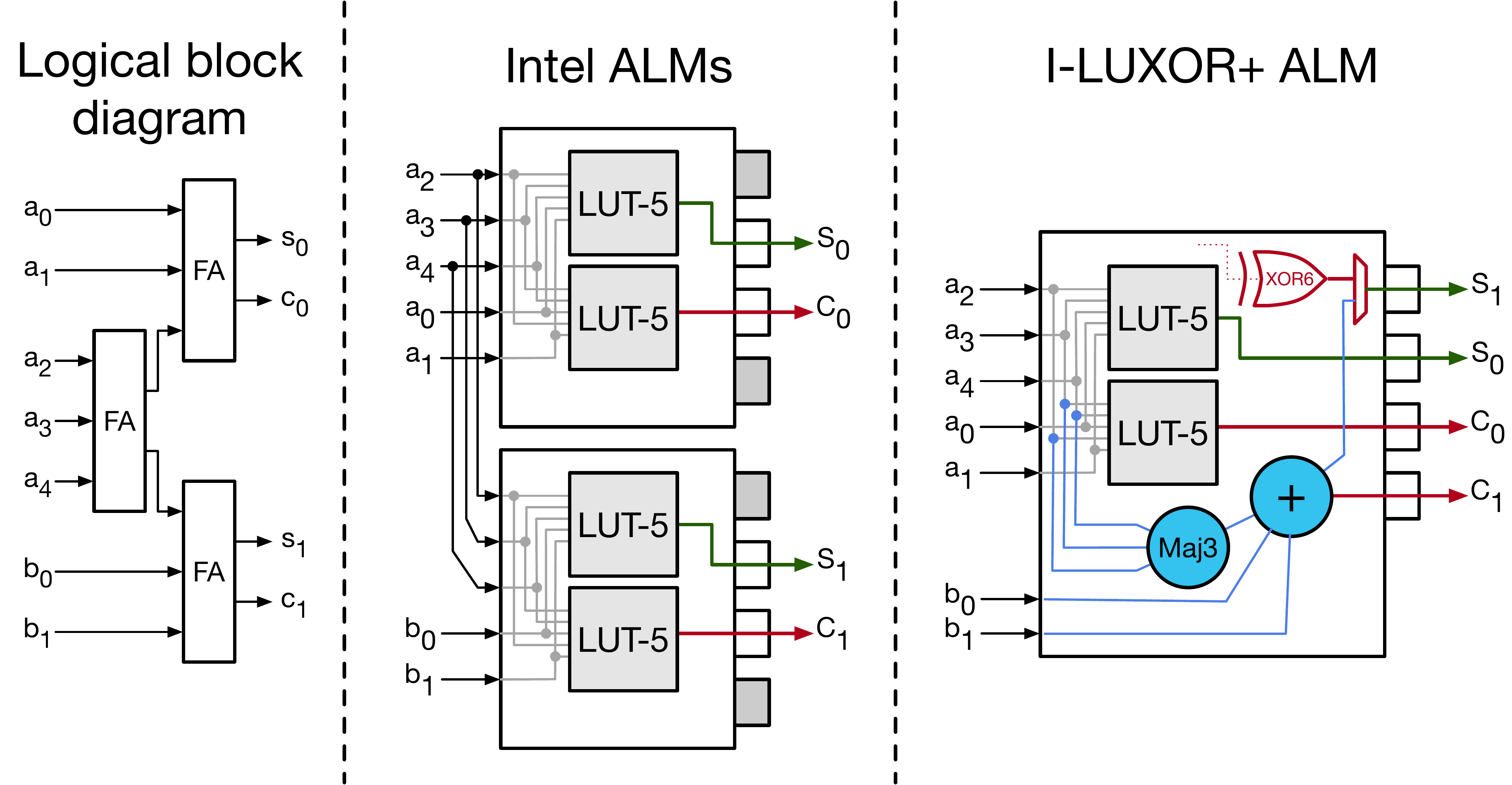}
    \vspace{-15pt}
    \caption{Efficient implementation of C25:121 GPC}
    \label{ramin_C25121.fig}
\end{figure}

Note that in Figure~\ref{dist_gpcs.fig}, the C25:121 GPC, originally suggested in~\cite{DBLP:journals/corr/abs-1806-08095}, is also a very efficient. Figure~\ref{ramin_C25121.fig} shows that it can be implemented using two sets of two 5-shared-input functions, occupying 2 ALMs. I-LUXOR+ introduces a majority circuit and full-adder to the ALM datapath, called MajFA (blue in Figure~\ref{intel_alm_mod.fig}), to explicitly implement S1 and C1 while S0 and C0 can be implemented in parallel with two 5-input LUT which shares the inputs in a ALM. This modification captures C25:121 in a single ALM instead of two. In summary, I-LUXOR+ reduces the cost of two highly used GPCs, C6:111 and C25:121, by one LUT (33\% and 50\% respectively).

\section{ILP-based Compressor Tree Synthesis}~\label{se:ilp}

\begin{figure}[h]
    \centering
    \includegraphics[width=2.25in]{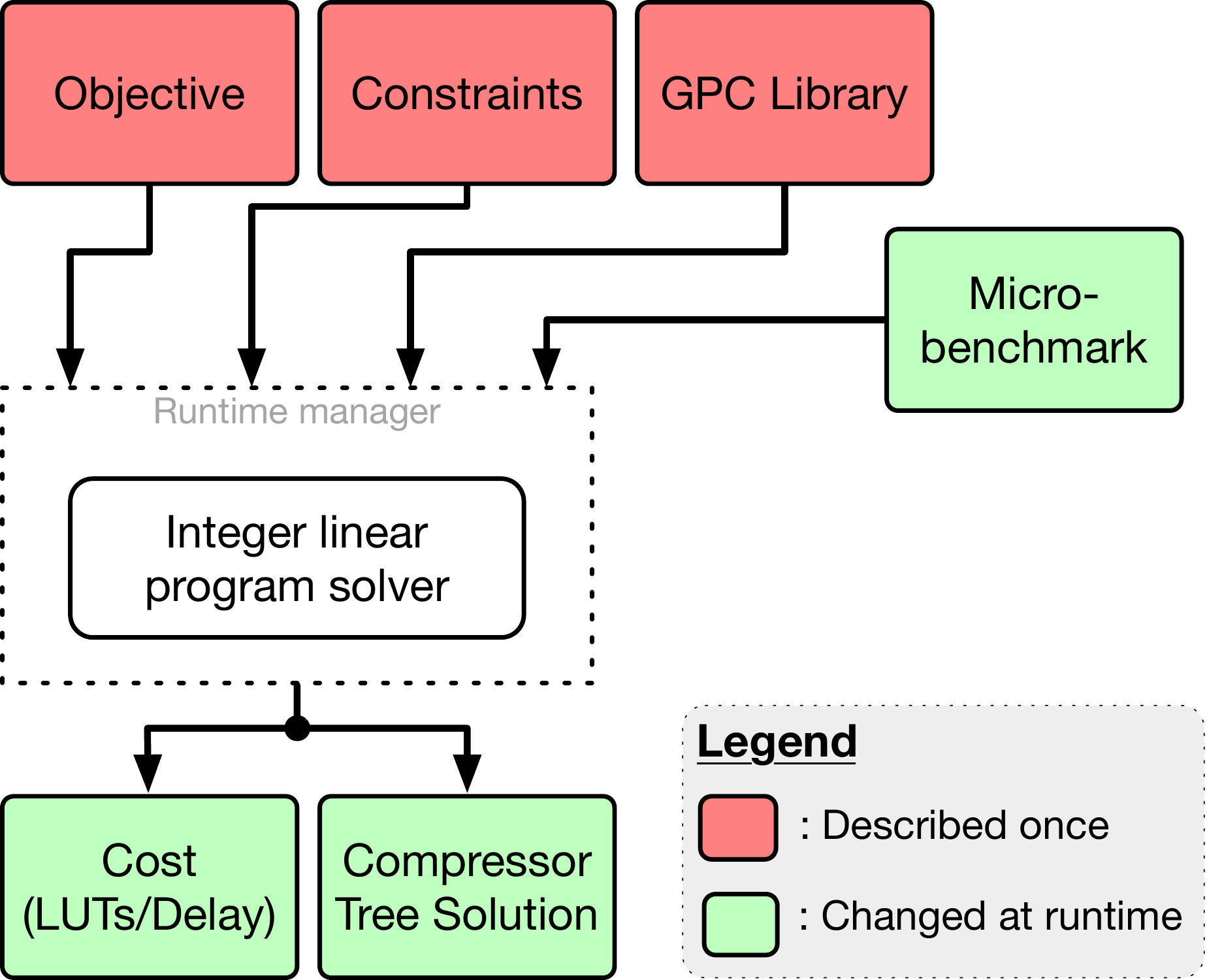}
    \vspace{-5pt}
    \caption{Flowchart of ILP-based compressor tree synthesis}
    \label{ilp_flow.fig}
\end{figure}


Many commonly used arithmetic operations such as multiplications, multiply-add, or digital filters can be expressed compactly as compressor tree hardware implementations. However, realizing efficient compressor trees is a non-trivial task that typically requires software automation. Methods to do efficient compressor tree synthesis include heuristics-guided search~\cite{DBLP:journals/corr/abs-1806-08095,oklobdzija1996method,stelling1998optimal}, integer linear programs (ILP)~\cite{DBLP:conf/date/VermaI07,DBLP:conf/fpl/KummZ14}, or hybrid approaches~\cite{kumm2018advanced}.
We opt for the ILP method in this work, and use ideas from~\cite{DBLP:conf/fpl/KummZ14}~and~\cite{DBLP:journals/corr/abs-1806-08095} as inspiration. Our goal is to quantify the effect of our proposed LUXOR/LUXOR+ modifications on efficient compressor tree synthesis for commonly-used arithmetic operations in modern applications. Figure~\ref{ilp_flow.fig} encapsulates the workflow of our ILP formulation, and we detail each building block shown in the figure. Table~\ref{ilp_legend.tbl} serves as a reference for all the variables used in this section. Note that, for clarity, all variable names in Table~\ref{ilp_legend.tbl} are local to this section, and should not be confused with nomenclature in other sections.





\begin{table}
	\centering
    \caption{Variables used in the ILP model}
	\label{ilp_legend.tbl}

	\newcolumntype{M}[1]{>{\centering\arraybackslash}m{#1}}
	\newcolumntype{L}[1]{>{\arraybackslash}m{#1}}
	
    \begin{tabular}{M{7mm}L{70mm}}
		\toprule
        \multicolumn{1}{c}{\bf Var} & \multicolumn{1}{c}{\bf Description} \\ 
        \midrule
        St & Number of stages in model \\
        C & Maximum number of columns in model \\
        $X_c$ & Number of bits in column $c$ of benchmark \\
        T & Total number of compressors used \\
        $I_t$ & Total number of columns consumed by compressor $t$ \\
        $V_t$ & Cost (in LUTs) of compressor $t$ \\
        $M_{t,c}$ & Number of bits consumed by compressor $t$ in column $c$ \\
        $O_t$ & Total number of columns output by compressor $t$ \\
        $K_{t,c}$ & Number of bits output by compressor $t$ in column $c$ \\
        $N_{s,c}$ & Number of bits in stage $s$ of column $c$ \\
        $C_{s,c}$ & Number of carry-bits in stage $s$ of column $c$ \\
        $R_{s,t,c}$ & Number of compressor $t$ used in column $c$ of stage $s$ \\
		\bottomrule
	\end{tabular}
\end{table}

\subsection{Objective}

There are two key metrics that quantify the effectiveness of a compressor tree implementation on FPGAs: area utilization in LUTs and the critical path delay, which is strongly correlated to the number of stages in the compressor tree. Hence, the objective function to an ILP program should be described in a way that minimizes these two metrics for each input micro-benchmark. To minimize the area cost, the objective function can be written as follows:


\begin{align*}
    & \min \sum_{s=0}^{St-1} \sum_{c=0}^{C-1} \sum_{t=0}^{T-1} V_{t}R_{s,t,c}
\end{align*}

To model the number of stages in the objective function, the authors in~\cite{DBLP:conf/fpl/KummZ14} add the number of
stages (St) as a heuristic to the cost function. However, we found this optimization strategy to be slow for
difficult problems, and in some cases,  the solver returns a solution that takes more
stages than required. To tackle this issue, we design a runtime manager that improves the speed of the optimization process.

\subsection{Runtime manager and solver}
Instead of modeling St as a heuristic in the objective function, we rely on an iterative approach where we query the solver to find an optimal solution within a fixed maximum stage
limit, St$_{max}$. This limit is relaxed incrementally until a feasible solution is found. 
In practice, we found that the solver was able to determine
infeasibility within a few seconds, whilst being able to find a feasible integer solution
within a few minutes. This iterative
approach was also recently used by Kumm~et.~al.~\cite{kumm2018advanced} by combining the ILP optimality search with heuristics to guide the solver. We use the IBM CPLEX v12.9~\cite{cplex2009v12} ILP solver (under academic license), and design a Python3-based interface for the runtime manager using the PuLP package~\cite{Mitchell11pulp:a}.


\subsection{Constraints}


Since the the input stage captures the input shape of the benchmark, 
we set constraints on the input stage 
as follows:
\begin{align*}
    & N_{0,c} = X_c \hspace{0.2in}\text{for c = 0,1,2,...,C-1}
\end{align*}


For subsequent stages, there are two constraints required to guide the solver towards a feasible compressor tree
architecture, such that input/output requirements of each stage are met:

\begin{minipage}{1in}
\begin{align*}
    & \sum_{t=0}^{T-1} \sum_{c'=0}^{O_t-1} M_{t,c'} * R_{s-1,t,c-c'} \ge N_{s-1,c}\\
    & \sum_{t=0}^{T-1} \sum_{c'=0}^{I_t-1} K_{t,c'} * R_{s-1,t,c-c'} = N_{s,c}\\
\end{align*}
\end{minipage}\hspace{0.25in}
\begin{minipage}{1in}
    \text{for c = 0,1,2,...,C-1}\\
    \text{for s = 1,2,3,...,St-1}\\
\end{minipage}

\noindent
The first constraint ensures that all bits in each column of every stage are
used as inputs by compressors in the next stage. The second constraint ensures
that the number of bits produced by the compressors in the previous stage
matches the number of input bits in the following stage. Both these constraints
can also be found in~\cite{DBLP:conf/fpl/KummZ14}.


In each stage, the number of carry-bits in each column are computed in \eqref{eq:carry}, where the division by two is due to the  increase in the column's radix. 

\begin{equation}
    C_{s,c} = \floor*{\frac{C_{s,c-1} + N_{s,c-1}}{2}} \label{eq:carry}
\end{equation}

This can be formulated as an ILP constraint as follows:

\begin{minipage}{1in}
\begin{align*}
    & C_{s,c} + 0.999 \ge \frac{1}{2}(C_{s,c-1} + N_{s,c-1}) \\
    & C_{s,c} \le \frac{1}{2}(C_{s,c-1} + N_{s,c-1}) \\
    & C_{s,0} = 0 \\
\end{align*}
\end{minipage}\hspace{0.25in}
\begin{minipage}{1in}
    \text{for c = 0,1,2,...,C-1}\\
    \text{for s = 0,2,3,...,St-1}\\
\end{minipage}

\noindent
Note that the number of input carry-bits into the first column is always set to 0.


When solving the model iteratively, as described above, the constraints on the
final stage guide the solver to converge to the solution.
In~\cite{DBLP:journals/corr/abs-1806-08095}, the author proposes a novel ragged
carry-propagate architecture for the final accumulation stage for Xilinx
FPGAs. This architecture reduces the overall number of stages required, and
hence, we opt for this strategy on Xilinx FPGAs.
Unlike~\cite{DBLP:journals/corr/abs-1806-08095}, where the author uses a
heuristic solver, we model the ragged carry-propagate adder into our
model for the final stage as three constraints:

\begin{minipage}{1in}
\begin{align*}
    & N_{s,c} + C_{s,c} \le 5 \\
    & C_{s,c} \le 2 \\
    & N_{s,c} \le 4 \\
\end{align*}
\end{minipage}\hspace{0.25in}
\begin{minipage}{1in}
    \text{for c = 0,1,2,...,C-1 and s = St-1}\\
\end{minipage}

\noindent
Finally, since Intel FPGAs cannot benefit from the ragged carry-propagate
adder, we model the ILP constraints for Intel FPGAs as shown
in~\cite{DBLP:conf/fpl/KummZ14} :

\begin{minipage}{1.5in}
\begin{align*}
    & N_{s,c} \le 3 \\
\end{align*}
\end{minipage}%
\begin{minipage}{1.5in}
    for c = 0,1,2,...,C-1 and s = St-1
\end{minipage}

\noindent




\subsection{GPC/Compressor library}


\subsubsection{Xilinx compressor set}


When targeting Xilinx architectures for our baseline, we use the GPC/compressor set defined by Preu{\ss}er \cite{DBLP:journals/corr/abs-1806-08095}, who pruned a set from Kumm and Zipf~\cite{DBLP:conf/mbmv/KummZ14}. For our LUXOR experiments, we reduce the cost of C6:111 GPCs from 3 to 2 logic elements, as described in Section~\ref{sssec: Xlux+}. For LUXOR+, in addition to the smaller version of C6:111, we add all the new slice-based GPCs described in Table~\ref{table_Xilinx_new_slice_based_compressor} to our model. We denote these results as X-LUXOR and X-LUXOR+ respectively.



\subsubsection{Intel compressor set}

When targeting Intel architectures, our baseline compressor set is based on a GPC set proposed by Parandeh-Afshar et al. \cite{DBLP:journals/trets/Parandeh-AfsharNBI11}, augmented with the C25:121 compressor from \cite{DBLP:journals/corr/abs-1806-08095}. Since this GPC set is large, to minimise run-time of our ILP, we pruned this set using  the GPC selection approach and metric described by Preu{\ss}er \cite{DBLP:journals/corr/abs-1806-08095}.


Parandeh-Afshar~et~al.~\cite{DBLP:journals/trets/Parandeh-AfsharNBI11} have gathered a group of LUT-based and arithmetic-based GPCs for Intel architectures. In the first three line of Table~\ref{table_intel_GPCs}, we show the efficiency and compression metrics of our selected GPCs according to the $APD$ (Equation~\ref{equ_GPC_APD}) metric, which measures the efficiency of a GPC taking into account delay and resource usage. We also considered the delay itself, since some of the proposed GPCs, such as C7:111, offer slightly better $S$ (compression rates) but their reported delay is 3.5$\times$ greater. In addition, we included C3:11 and C25:121 in the baseline GPC set for Intel architecture.  

Similar to our X-LUXOR experiments with Xilinx architectures, we reduce the cost of C6:111 GPCs from 3 to 2 logic elements for I-LUXOR. For I-LUXOR+, as well as using the upgraded version of C6:111, we reduce the cost of C25:121 from  2 to 1 LE as described in Section~\ref{se:majfa}. 
We also comment that the effect of the I-LUXOR and I-LUXOR+ enhancements are highlighted by the metrics, as demonstrated by the last two rows of Table\ref{table_intel_GPCs}. Due to the lower logic element cost, the $APD$ of both GPCs show significant improvement.

\begin{table}
\setlength{\tabcolsep}{2pt}

\newcolumntype{M}[1]{>{\centering\arraybackslash}m{#1}}
\newcolumntype{L}[1]{>{\arraybackslash}m{#1}}

\caption{Comparison of different GPCs proposed in \cite{DBLP:journals/corr/abs-1806-08095} and new GPCs supported by I-LUXOR and I-LUXOR$+$}
\label{table_intel_GPCs}
\vspace{-7pt}
\centering
\begin{tabular}{|c|M{15mm}||c|c|c|c|c|}
    \hline
    \multicolumn{2}{|c||}{\bf GPCs} & {$\bm{S}$} & {$\bm{A}$} & {\bf Delay} & {\bf LUTs} & {\bf APD}\\
    \hline
    \hline
    \multirow{3}{*}{\cite{DBLP:journals/trets/Parandeh-AfsharNBI11}}&{C6:111} & {2} & {0.13} & {0.38} & {3}  & {7.9}\\
    \cline{2-7}
    {}&{C15:111} & {2} & {0} & {0.38} & {3} & {7.9} \\
    \cline{2-7}
    {}&{C23:111} & {1.67} & {0} & {0.38} & {2} & {5.3} \\
    \cline{2-7}
    \hline
    \hline
    {\cite{DBLP:journals/corr/abs-1806-08095}}&{C25:121} & {1.75} & {0} & {0.38} & 2 & {11.8}\\
    \hline
    \hline
    \multirow{2}{*}{\rotatebox{90}{\!\!Ours}}&{C6:111} & {2} & {0.13} & {0.39*} & {2} & {10.95*}\\
    \cline{2-7}
    {}&{C25:121} & {1.75} & {0} & {0.39*} & {1} & {21.9*}\\
    \hline
    \multicolumn{7}{L{80mm}}{\footnotesize *Area/delay overheads for I-LUXOR+ are included (Section~\ref{se:results}).}
\end{tabular}
\end{table}


\subsection{Micro-Benchmarks}

To evaluate the improvements of our proposed architectures, we use different basic operations that are commonly found in various domains in three categories: 1) Low-rank inputs including pop-count and two-column count (based on~\cite{DBLP:journals/corr/abs-1806-08095}, but with additional input sizes) 2) High-rank inputs including multi-addition \cite{DBLP:conf/fpl/KummZ14}, 3-MAC operation (described below), and a FIR-3 filter from~\cite{DBLP:journals/trets/Parandeh-AfsharNBI11}, and 3) BNN XnorPopcount operation for various input sizes, where the filter sizes are taken from the networks in \cite{DBLP:journals/corr/abs-1809-04570,DBLP:conf/nips/HubaraCSEB16}. These three categories highlight the benefits and limitations of LUXOR and LUXOR+ architectures, as the chosen operations appears in various applications, especially digital signal processing and neural networks which are the most important concerns of new FPGA architectures~\cite{boutros2019math,DBLP:conf/fccm/RasoulinezhadZW19}.

\subsubsection{3-MAC operation}
The 3--MAC operation is modeled according to the following equation:
\begin{equation} 
   \label{eq_APD}
   \begin{split}
   \text{3-MAC}_{N\PLH (N-bit)} = \sum_{i=0}^{2} A_{i(N-bits)}\times B_{i(N-bits)}
   \end{split}
\end{equation} 

Note that since there are 3 pairs of inputs, instead of computing partial products then and summing their results, we can select partial products of the same rank and perform a primary compression. The cost of this step is included in our result. The resulting tree forms the input to the compressor. We repeat this for different input widths ($N$).


\section{Results}
\label{se:results}
In this section we present results from experiments undertaken to evaluate the performance of the LUXOR and LUXOR+ architectural enhancements. 

\subsection{ASIC Modeling: Delay and Area Overheads}

We model state-of-the-art Intel Stratix-10 ALM unit~\cite{ug_s10_lab}, and Xilinx UltraScale+ slice~\cite{ug574,ug474} according to their respective data sheet descriptions. For the ASIC metric analysis, we synthesize our Verilog models using SMIC 65-nm technology standard cell by Synopsis Design Compiler 2013.12. Post synthesis results are reported and the synthesis strategy was set to ``Timing Optimization'' since it usually leads to a better $Area \times Delay$ product. We note that while our approach to estimating area and delay overheads using standard cells may differ slightly from an commercial full custom layout, in either case the overhead is minimal.

Table~\ref{table_ASIC_ALM} gives the post-synthesis area and timing results for the Intel baseline, I-LUXOR, Intel+MajFA and I-LUXOR+ modifications to the ALM. From the table, it can be seen that the delay increase of I-LUXOR is about $1\%$ while the area increase is less than $0.5\%$. This demonstrates that there is little overhead associated with adding a 6-input XOR gate to the ALM unit. In contrast, adding MajFA circuits will increase the area and delay by $2\%$ and $5\%$ respectively (see description in Section~\ref{se:majfa}). The full I-LUXOR+ implementation, has $3\%$ and $5\%$ delay and area overhead respectively. We believe that the unexpectedly large increase in area compared to the individual effect of each modification arises from the performance-driven synthesis optimization. For measuring the critical path, we removed the multiplexers connecting the ALM's outputs to its input, and thus it is measured from: an input, through a LUT and two-coupled full adders (FAs) to an output multiplexer.

\begin{table}[t]
\setlength{\tabcolsep}{4pt}
\caption{ASIC results for the Intel Stratix-10 ALM architecture}
\label{table_ASIC_ALM}
\vspace{-7pt}
\centering
\begin{tabular}{|c|c|c|c|c|c|}
    \cline{3-6}
    \multicolumn{2}{c|}{} & {Intel} & {I-LUXOR} & {Intel+MajFA} & {I-LUXOR+}\\
    \hline
    \multirow{2}{*}{Area} & {$um^2$} & {1680} & {1687} & {1715} & {1767}\\
    \cline{2-6}
    {} & {ratio} & {1} & {1.00} & {1.02} & {1.05}\\
    \hline
    \multirow{2}{*}{Delay} & {$ns$} & {1.42} & {1.44} & {1.49} & {1.46}\\
    \cline{2-6}
    {} & {ratio} & {1} & {1.01} & {1.05} & {1.03}\\
    \hline
\end{tabular}
\end{table}

In a Xilinx slice, the critical path is from an input, passing through the first LUT (A) and four carry-chain circuits, and ending with the last output multiplexer. This path is also the critical path after applying  LUXOR(+) for both architectures.

The synthesized Xilinx baseline slice model has an area of 6045~$um^2$. We compare the reported critical path with that from the Virtex-5 datasheet, which was a device that was also manufactured with the similar 65~$nm$ process. Reference~\cite{DS202} reports the critical path from an input, through four carry circuits to the output ($T_{ITO}$) as 0.67, 0.77, or 0.90-$ns$ for three different speed grades. Comparing these values with our value of 0.84~$ns$ from Table~\ref{table_Xilinx_slice}, consistency with our synthesis results was verified. 
The same table shows that X-LUXOR has similar area utilization and a 6\% increase in delay, while X-LUXOR+ has 6\% area and 9\% delay overheads. 

\begin{table}[t]
\setlength{\tabcolsep}{4pt}
\caption{ASIC results for the Xilinx UltraScale+ slice architecture}
\label{table_Xilinx_slice}
\vspace{-7pt}
\centering
\begin{tabular}{|c|c|c|c|c|}
    \cline{3-5}
    \multicolumn{2}{c|}{} & {Xilinx} & {X-LUXOR} & {X-LUXOR+}\\
    \hline
    \multirow{2}{*}{Area} & {$um^2$} & {6045} &{6002} & {6361}\\
    \cline{2-5}
    {} & {ratio} & {1.00} & {0.99} & {1.06}\\
    \hline
    \multirow{2}{*}{Delay} & {$ns$} & {0.84} & {0.89} & {0.92}\\
    \cline{2-5}
    {} & {ratio} & {1.00} & {1.06} & {1.09}\\
    \hline
\end{tabular}
\end{table}

Since the routing delay strongly contributes to the total delay, the LUXOR(+) delay advantages are diluted in practice. Although the X-LUXOR+ overheads are notable, because of the significant resource and performance benefits, new trade-offs are offered. For example, partially upgrading the LEs to LUXOR(+) architectures is another option. Also, with more effort in layout and buffer sizing, area and delay overheads can be recovered/balanced.

At a higher level of abstraction, LUXOR(+) does not require any I/O scheme modifications. However, they increase the logic implementation density leading to higher connectivity per LE/ALM. Thus, routing limitations may slow down LUXOR(+) enhancements. LUXOR(+) adds to the input load which also slows down the LE. This was not measured directly but taken into account in the LE measurements.


\subsection{Benchmark Performance}

The effect of our ILP approach on resource utilization in logic elements is affected by the choice of primitives in the primary stage (if applicable), compression tree stages, and the last stage (final ternary adder in Intel or the equivalent relaxed ternary adder for  Xilinx architectures as proposed in~\cite{DBLP:journals/corr/abs-1806-08095}). Table~\ref{table_ilp_tomas} compares our technique with that of~\cite{DBLP:journals/corr/abs-1806-08095} for X-LUXOR and X-LUXOR+ where test cases are popcount and double column popcount operations indicated respectively by S and D, concatenated with input size. 
As can be seen in the baseline column, our ILP approach uses fewer logic elements (LEs) and stages for all benchmark problems compared with the heuristic approach, since an optimal solution is found. While the X-LUXOR enhancement significantly reduces the number of LEs compared with the baseline, X-LUXOR+ achieves a further reduction in the number of stages.

\begin{table}[t]
\setlength{\tabcolsep}{2.75pt}
\caption{A comparison of solutions from our ILP-based synthesis compared with those reported in~\cite{DBLP:journals/corr/abs-1806-08095}}
\label{table_ilp_tomas}
\vspace{-7pt}
\centering
\begin{tabular}{|c|c|c|c|c|c|c|c|c|}
    \hline
    \multirow{2}{*}{Test} & \multicolumn{2}{c|}{\multirow{2}{*}{$^1$H1/H2/H3\cite{DBLP:journals/corr/abs-1806-08095}}} & \multicolumn{6}{c|}{Our ILP Solver}\\
    \cline{4-9}
    \multirow{2}{*}{cases} & \multicolumn{2}{c|}{} & \multicolumn{2}{c|}{Baseline}& \multicolumn{2}{c|}{X-LUXOR} & \multicolumn{2}{c|}{X-LUXOR+} \\
    \cline{2-9}
    {} & {$^2$LE} & {$^2$Stage} & {LE} & {Stage} & {LE} & {Stage} & {LE} & {Stage} \\
    \hline 
    {S128} & {101/102/101} & {4/3/4} & {100} & {3} & {79} & {3} & {78} & {3} \\
    \hline
    {S256} & {209/209/209} & {4/4/4} & {195} & {4} & {159} & {4} & {154} & {4} \\
    \hline
    {S512} & {418/422/418} & {5/5/5} & {380} & {5} & {319} & {5} & {312} & {4} \\
    \hline
    {D128} & {178/205/178} & {5/4/5} & {168} & {4} & {156} & {4} & {150} & {4} \\
    \hline
    {D256} & {360/417/360} & {6/5/6} & {328} & {5} & {315} & {5} & {298} & {4} \\
    \hline
    {D512} & {721/839/721} & {7/6/7} & {709} & {5} & {631} & {5} & {586} & {5} \\
    \hline
    \multicolumn{9}{l}{$^1$\footnotesize Heuristics used in~\cite{DBLP:journals/corr/abs-1806-08095}: Efficiency/Strength/Product, reported in that order.} \\
    \multicolumn{9}{l}{$^2$\footnotesize LE = logic elements (LUTs), Stage = \# of compressor tree stages} \\
\end{tabular}
\end{table}

\begin{figure*}[th]
    \centering
    \includegraphics[width=\textwidth]{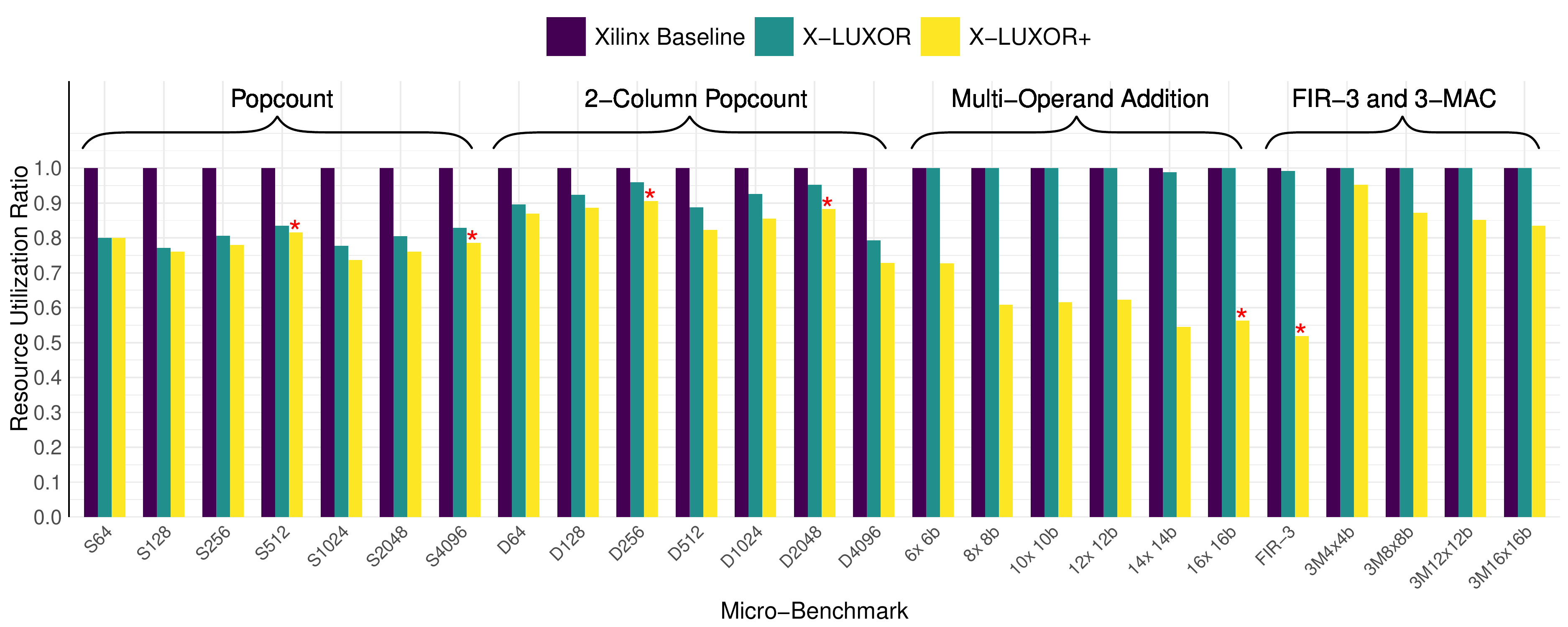}
    \vspace{-15pt}
    \caption{Resource reduction on Xilinx UltraScale+, X-LUXOR, and X-LUXOR+ architectures for various micro-benchmarks. The {\color{red}*} indicates that the proposed solution required one less logic stage in the compression tree.}
    \label{xilinx_impact.fig}
\end{figure*}

Figure~\ref{xilinx_impact.fig} shows the savings in LEs for Xilinx architectures over a larger micro-benchmark set, with the red star also indicating a reduction in number of stages by one. For low-rank inputs (i.e. popcount and two-column popcount), the C6:111 and C25:121 compressors are heavily used. X-LUXOR improves the resource efficiency of C6:111 implementations and achieves the best savings for the 1024-input popcount problem at 22\% reduction. Less improvement is seen for two-column popcount, as in the first stage, C25:121 has better arithmetic slack ($A$) while offering the same efficiency. This observation was also made in~\cite{DBLP:journals/corr/abs-1809-04570}. X-LUXOR+ offers a new set of the state of the art compression rate and compression efficiency. On average, X-LUXOR+ can reduce area utilization on the low-rank input popcount and two-column popcount benchmarks by 22\% and 15\% respectively.

For the high-rank benchmarks (multi-operand addition, FIR-3 and 3-MAC), the inputs are wide enough to benefit from the slice based GPCs. The C6:111 compressor is not significantly utilized. However, X-LUXOR+ offers higher compression rates and hence achieves 39\% and 18\% improvement in multi-addition and 3-MAC benchmarks respectively and in some cases the required number of stages is also reduced by one. 

\begin{figure*}[th]
    \centering
    \includegraphics[width=\textwidth]{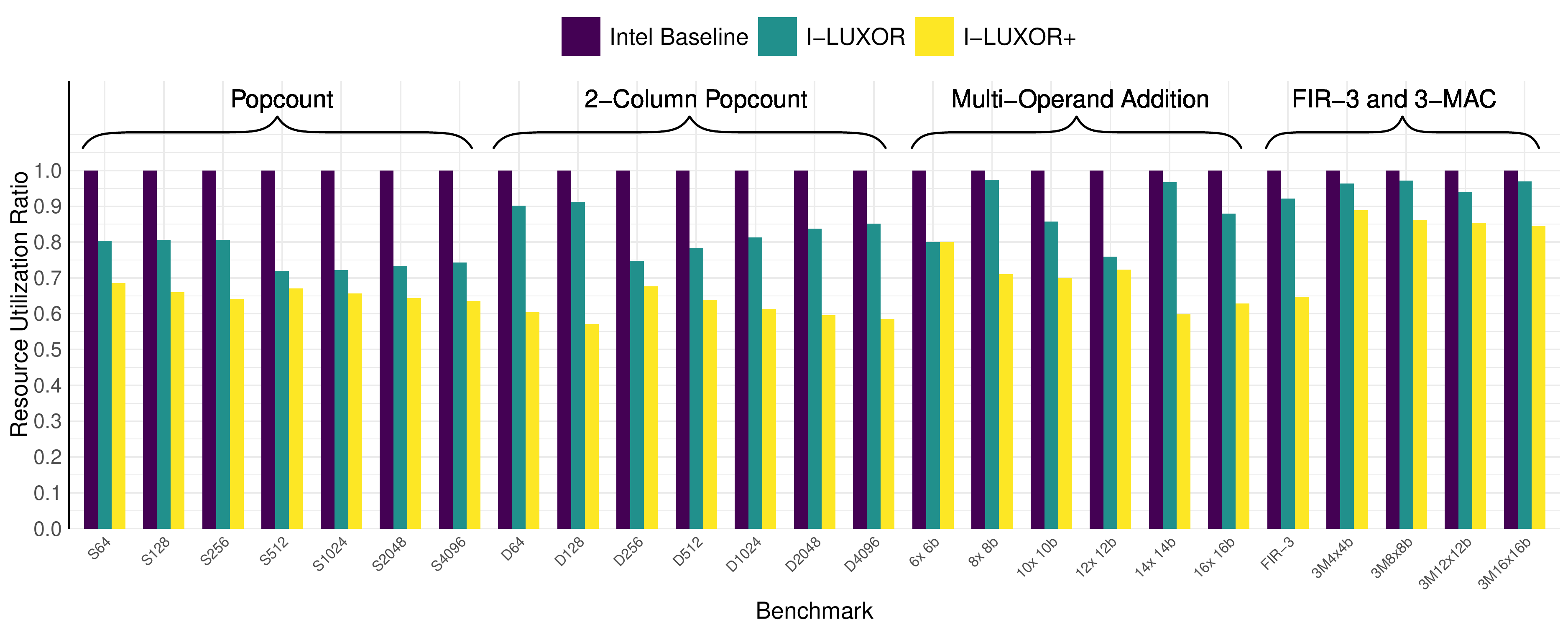}
    \vspace{-15pt}
    \caption{Resource reduction on Intel Stratix-10, I-LUXOR, and I-LUXOR+ architecture for our selected micro-benchmarks.}
    \label{intel_impact.fig}
\end{figure*}

Figure~\ref{intel_impact.fig} shows the same result for Intel I-LUXOR, and I-LUXOR+ architectures. More dramatic resource savings are apparent over Xilinx, particularly for low-rank problems using I-LUXOR+. Since I-LUXOR and I-LUXOR+ do not present new compressors, no reduction in number of stages is achieved. However, because the baseline offers no wide GPC, the resource reduction of I-LUXOR is more significant (averaging 24\% and 17\% for popcount and double popcount). I-LUXOR+ offers an enhanced C25:121 GPC which is the most efficient GPC for the Intel architecture. This leads to 35\% and 39\% resource savings for popcounting and two-column counting.

\subsection{Performance on BNNs}
Binarized neural networks offer a new challenge for FPGA architectures as 1-bit multiply-accumulate operations require XNOR and popcount operations to be efficient. As explained in Section~\ref{se:bnn} the first computation stage (Multiplication) should be merged with the early compression circuits, leading to an efficient implementation (as illustrated in Figure~\ref{bnn_idea.fig}(a)). If the number of input pairs is $N$, $N/3$ fused units are required in the primary stage. LUXOR can implement this fused computation using a single LE rather than two LEs in the baseline architectures leading to $N/3$ fewer LE utilization. In addition, after implementation of the primary stage, a two-column counting problem with the height of $N/3$ is encountered.  

\begin{figure}[th]
    \centering
    \includegraphics[width=\linewidth]{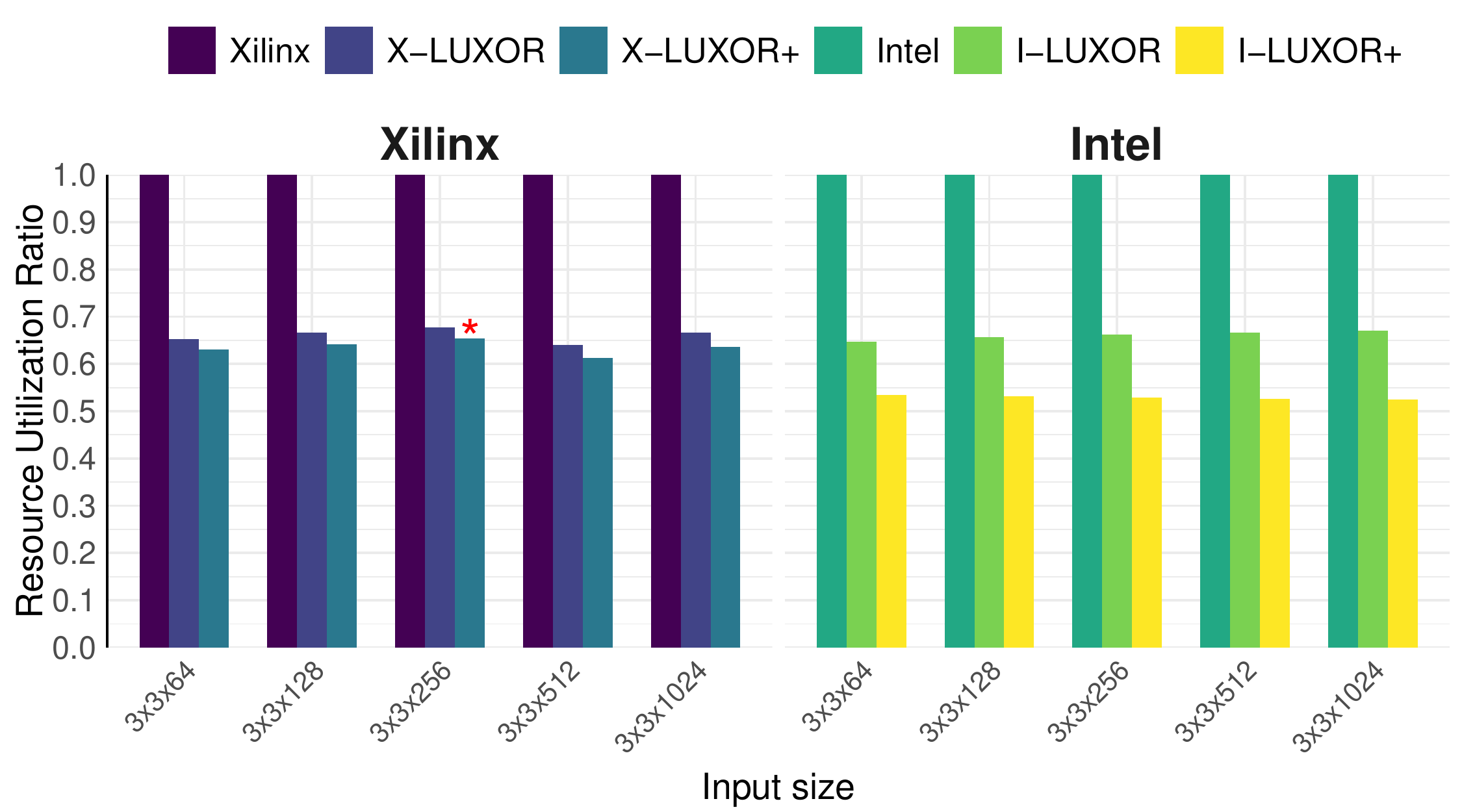}
    \vspace{-15pt}
    \caption{XnorPopcount micro-benchmarks found in BNN Convolution Layers in~\cite{umuroglu2017finn,DBLP:conf/nips/HubaraCSEB16}}
    \label{fig_plot_bnn}
\end{figure}

As shown in Figure~\ref{fig_plot_bnn}, these two optimizations lead to almost the same 34\% resource reduction for LUXOR modification on both Xilinx and Intel architectures. Moreover, as described before, X-LUXOR+ cannot reduce the number of LEs significantly for low-rank inputs, and hence, the best area savings for BNNs plateaus at 37\%. In the case when the input size is 3$\times$3$\times$256, the number of stages is reduced by one, which would give us a significant improvement in delay. In comparison, I-LUXOR+ reduces the number of LEs significantly at an average of 47\%, but without reducing the number of stages.

\section{Conclusion}
\label{se:conclusion}
This paper has discussed several low-cost FPGA logic cell modifications that can lead to significantly improved performance GPCs and the XnorPopcount operation. We then added these primitives to a set of state-of-the-art compressor tree primitives (adders, GPCs, and compressors) described in literature and built an ILP model for finding optimal solutions for FPGA-based compressor tree implementations for both Xilinx and Intel FPGAs. Using this ILP, we were able to show that our modifications lead to substantial performance gains. LUXOR is a vendor-agnostic modification, which augments each logic cell datapath with a dedicated 6-input XOR circuit, reduces the cost of the commonly used (C$6\text{:}111$) GPC from 3 LEs to just 2 and enables efficient XnorPopcount implementations.
Over a benchmark set, this reduces the logic utilization cost of compressor trees by up to 36\% (average 12--19\%) on both Intel and Xilinx FPGAs, with a silicon area overhead of $<$0.5\%. The architectural re-design is taken a step further with LUXOR+, which proposes carefully-crafted vendor-specific modifications. LUXOR+ requires an additional 3--6\% silicon area, can improve our micro-benchmark results to up to 48\% (average 26--34\%).

\bibliography{library}{}

\begin{thebibliography}{10}

\bibitem{DBLP:journals/corr/abs-1809-04570}
Michaela Blott, Thomas~B. Preu{\ss}er, Nicholas~J. Fraser, Giulio Gambardella,
  Kenneth O'Brien, Yaman Umuroglu, Miriam Leeser, and Kees~A. Vissers.
\newblock {FINN-\emph{R}}: An end-to-end deep-learning framework for fast
  exploration of quantized neural networks.
\newblock {\em {TRETS}}, 11(3):16:1--16:23, 2018.

\bibitem{boutros2019math}
Andrew Boutros, Mohamed Eldafrawy, Sadegh Yazdanshenas, and Vaughn Betz.
\newblock Math doesn't have to be hard: Logic block architectures to enhance
  low-precision multiply-accumulate on {FPGAs}.
\newblock In {\em Proceedings of the 2019 ACM/SIGDA International Symposium on
  Field-Programmable Gate Arrays {FPGA}}, pages 94--103. ACM, 2019.

\bibitem{cplex2009v12}
IBM~ILOG CPLEX.
\newblock V12. 1: User\textquoteright s manual for cplex.
\newblock {\em International Business Machines Corporation}, 46(53):157, 2009.

\bibitem{dadda1965some}
Luigi Dadda.
\newblock Some schemes for parallel multipliers.
\newblock {\em Alta frequenza}, 34:349--356, 1965.

\bibitem{earle65}
J.~G. Earle.
\newblock Latched carry-save adder.
\newblock {\em IBM Tech. Disc. Bull.}, 7(10):909--910, 1965.

\bibitem{ercegovac2004digital}
Milos~D Ercegovac and Tomas Lang.
\newblock {\em Digital arithmetic}.
\newblock Elsevier, 2004.

\bibitem{DBLP:conf/nips/HubaraCSEB16}
Itay Hubara, Matthieu Courbariaux, Daniel Soudry, Ran El{-}Yaniv, and Yoshua
  Bengio.
\newblock Binarized neural networks.
\newblock In {\em Advances in Neural Information Processing Systems 29: Annual
  Conference on Neural Information Processing Systems}, pages 4107--4115, 2016.

\bibitem{ug_s10_lab}
Intel.
\newblock {UG-S10LAB} {Intel\textregistered Stratix\textregistered 10 Logic
  Array Blocks and Adaptive Logic Modules User Guide}, 9 2018.

\bibitem{DBLP:conf/fpt/KimLA18}
Jin~Hee Kim, Jongeun Lee, and Jason Anderson.
\newblock {FPGA} architecture enhancements for efficient {BNN} implementation.
\newblock In {\em International Conference on Field-Programmable Technology,
  {FPT}}, pages 214--221, 2018.

\bibitem{kumm2018advanced}
Martin Kumm and Johannes Kappauf.
\newblock Advanced compressor tree synthesis for {FPGAs}.
\newblock {\em IEEE Transactions on Computers}, 67(8):1078--1091, 2018.

\bibitem{DBLP:conf/mbmv/KummZ14}
Martin Kumm and Peter Zipf.
\newblock Efficient high speed compression trees on xilinx {FPGAs}.
\newblock In {\em Methoden und Beschreibungssprachen zur Modellierung und
  Verifikation von Schaltungen und Systemen, B{\"{o}}blingen, Germany}, pages
  171--182, 2014.

\bibitem{DBLP:conf/fpl/KummZ14}
Martin Kumm and Peter Zipf.
\newblock Pipelined compressor tree optimization using integer linear
  programming.
\newblock In {\em 24th International Conference on Field Programmable Logic and
  Applications, {FPL}}, pages 1--8, 2014.

\bibitem{liang2018fp}
Shuang Liang, Shouyi Yin, Leibo Liu, Wayne Luk, and Shaojun Wei.
\newblock {FP-BNN}: Binarized neural network on {FPGA}.
\newblock {\em Neurocomputing}, 275:1072--1086, 2018.

\bibitem{meo1975arithmetic}
Angelo~Raffaele Meo.
\newblock Arithmetic networks and their minimization using a line of elementary
  units.
\newblock {\em IEEE Transactions on Computers}, 100(3):258--280, 1975.

\bibitem{Mitchell11pulp:a}
Stuart Mitchell, Stuart~Mitchell Consulting, and Iain Dunning.
\newblock Pulp: A linear programming toolkit for python, 2011.

\bibitem{oklobdzija1996method}
Vojin~G. Oklobdzija, David Villeger, and Simon~S. Liu.
\newblock A method for speed optimized partial product reduction and generation
  of fast parallel multipliers using an algorithmic approach.
\newblock {\em IEEE Transactions on computers}, 45(3):294--306, 1996.

\bibitem{parandeh2008efficient}
Hadi Parandeh-Afshar, Philip Brisk, and Paolo Ienne.
\newblock Efficient synthesis of compressor trees on {FPGAs}.
\newblock In {\em 2008 Asia and South Pacific Design Automation Conference},
  pages 138--143. IEEE, 2008.

\bibitem{DBLP:conf/fpl/Parandeh-AfsharBI09}
Hadi Parandeh{-}Afshar, Philip Brisk, and Paolo Ienne.
\newblock Exploiting fast carry-chains of {FPGAs} for designing compressor
  trees.
\newblock In {\em 19th International Conference on Field Programmable Logic and
  Applications, {FPL}}, pages 242--249, 2009.

\bibitem{DBLP:journals/trets/Parandeh-AfsharBI09}
Hadi Parandeh{-}Afshar, Philip Brisk, and Paolo Ienne.
\newblock An {FPGA} logic cell and carry chain configurable as a 6:2 or 7:2
  compressor.
\newblock {\em {TRETS}}, 2(3):19:1--19:42, 2009.

\bibitem{DBLP:journals/trets/Parandeh-AfsharNBI11}
Hadi Parandeh{-}Afshar, Arkosnato Neogy, Philip Brisk, and Paolo Ienne.
\newblock Compressor tree synthesis on commercial high-performance {FPGAs}.
\newblock {\em {TRETS}}, 4(4):39:1--39:19, 2011.

\bibitem{parandeh2009improving}
Hadi Parandeh-Afshar, Ajay~Kumar Verma, Philip Brisk, and Paolo Ienne.
\newblock Improving fpga performance for carry-save arithmetic.
\newblock {\em IEEE transactions on very large scale integration (VLSI)
  systems}, 18(4):578--590, 2009.

\bibitem{DBLP:journals/corr/abs-1806-08095}
Thomas~B. Preu{\ss}er.
\newblock Generic and universal parallel matrix summation with a flexible
  compression goal for xilinx fpgas.
\newblock In {\em 27th International Conference on Field Programmable Logic and
  Applications, {FPL}}, pages 1--7, 2017.

\bibitem{DBLP:conf/fccm/RasoulinezhadZW19}
SeyedRamin Rasoulinezhad, Hao Zhou, Lingli Wang, and Philip H.~W. Leong.
\newblock {PIR-DSP:} an {FPGA} {DSP} block architecture for multi-precision
  deep neural networks.
\newblock In {\em 27th {IEEE} Annual International Symposium on
  Field-Programmable Custom Computing Machines, {FCCM}}, pages 35--44, 2019.

\bibitem{rastegari2016xnor}
Mohammad Rastegari, Vicente Ordonez, Joseph Redmon, and Ali Farhadi.
\newblock {XNOR-Net}: Imagenet classification using binary convolutional neural
  networks.
\newblock In {\em European Conference on Computer Vision - {ECCV}}, pages
  525--542, 2016.

\bibitem{stelling1998optimal}
Paul~F Stelling, Charles~U Martel, Vojin~G Oklobdzija, and R~Ravi.
\newblock {Optimal circuits for parallel multipliers}.
\newblock {\em IEEE Transactions on Computers}, 47(3):273--285, 1998.

\bibitem{swartzlander1973parallel}
Earl~E Swartzlander.
\newblock Parallel counters.
\newblock {\em IEEE Transactions on computers}, 100(11):1021--1024, 1973.

\bibitem{umuroglu2017finn}
Yaman Umuroglu, Nicholas~J Fraser, Giulio Gambardella, Michaela Blott, Philip
  Leong, Magnus Jahre, and Kees Vissers.
\newblock {FINN: A framework for fast, scalable binarized neural network
  inference}.
\newblock In {\em Proceedings of ACM/SIGDA International Symposium on
  Field-Programmable Gate Arrays {FPGA}}, pages 65--74. ACM, 2017.

\bibitem{verma2008data}
Ajay~K Verma, Philip Brisk, and Paolo Ienne.
\newblock Data-flow transformations to maximize the use of carry-save
  representation in arithmetic circuits.
\newblock {\em IEEE Transactions on Computer-Aided Design of Integrated
  Circuits and Systems}, 27(10):1761--1774, 2008.

\bibitem{DBLP:conf/date/VermaI07}
Ajay~K. Verma and Paolo Ienne.
\newblock Automatic synthesis of compressor trees: reevaluating large counters.
\newblock In {\em 2007 Design, Automation and Test in Europe Conference and
  Exposition, {DATE} 2007, Nice, France, April 16-20, 2007}, pages 443--448,
  2007.

\bibitem{wallace1964suggestion}
Christopher~S. Wallace.
\newblock A suggestion for a fast multiplier.
\newblock {\em {IEEE} Trans. Electronic Computers}, 13(1):14--17, 1964.

\bibitem{DS202}
Xilinx.
\newblock {DS202 (v5.5)} {Virtex-5 FPGA Data Sheet:DC and Switching
  Characteristics}, 6 2016.

\bibitem{ug474}
Xilinx.
\newblock {UG474} {7 Series FPGAs Configurable Logic Block}, 9 2016.

\bibitem{ug574}
Xilinx.
\newblock {UG574} {UltraScale Architecture Configurable Logic Block}, 2 2017.

\bibitem{Zhao:2017:ABC:3020078.3021741}
Ritchie Zhao, Weinan Song, Wentao Zhang, Tianwei Xing, Jeng{-}Hau Lin, Mani~B.
  Srivastava, Rajesh Gupta, and Zhiru Zhang.
\newblock Accelerating binarized convolutional neural networks with
  software-programmable {FPGA}s.
\newblock In {\em Proceedings of the 2017 {ACM/SIGDA} International Symposium
  on Field-Programmable Gate Arrays, {FPGA}}, pages 15--24, 2017.

\end{thebibliography}
\bibliographystyle{plain}

\end{document}